\documentclass[10pt,twocolumn,aps,prd,preprintnumbers,showpacs,
superscriptaddress,nofootinbib,amsmath,amssymb,floats,floatfix,showkeys,
notitlepage,longbibliography]{revtex4-1}
\usepackage{graphicx}
\pdfoutput=1
\usepackage{bm}
\usepackage{amsmath,amssymb,amsfonts}
\usepackage{mathrsfs}
\usepackage[table]{xcolor}
\usepackage{colortbl}
\usepackage{longtable}
\usepackage{needspace}
\usepackage{array}
\usepackage{booktabs}
\usepackage{multirow}
\usepackage{orcidlink}
\usepackage{hyperref}
\hypersetup{colorlinks=true,linkcolor=blue,urlcolor=blue,citecolor=blue}
\numberwithin{equation}{section}

\begin{document}
\title{Spin-dependent topological structure of black-hole thermodynamics}

\author{Erdem Sucu \orcidlink{0009-0000-3619-1492}}
\email{23600348@emu.edu.tr}
\affiliation{Physics Department, Eastern Mediterranean University, 99628,
Famagusta, Cyprus}

\author{\.{I}zzet Sakall{\i} \orcidlink{0000-0001-7827-9476}}
\email{izzet.sakalli@emu.edu.tr}
\affiliation{Physics Department, Eastern Mediterranean University, 99628,
Famagusta, Cyprus}

\author{Yusuf Sucu \orcidlink{0000-0001-7874-0146}}
\email{ysucu@akdeniz.edu.tr}
\affiliation{Department of Physics, Faculty of Science, Akdeniz University, 07058
Antalya, T\"urkiye}

\author{Emmanuel N. Saridakis \orcidlink{0000-0003-1500-0874}}
\email{msaridak@noa.gr}
\affiliation{Institute for Astronomy, Astrophysics, Space Applications and Remote
Sensing, National Observatory of Athens, 15236 Penteli, Greece}
\affiliation{Departamento de Matem\'{a}ticas, Universidad Cat\'{o}lica del Norte,
Avda. Angamos 0610, Casilla 1280 Antofagasta, Chile}
\affiliation{Department of Astronomy, School of Physical Sciences, University of
Science and Technology of China, Hefei 230026, P.R. China}

\begin{abstract}
Thermodynamic topology classifies black-hole equilibrium states through
topological defects of an off-shell thermodynamic field, while
quantum-corrected Hawking radiation can retain information about the quantum
numbers of the emitted particle. Whether these two structures are connected
has remained largely unexplored. In this work we investigate this question for
a massive-gravity AdS black hole by combining arbitrary-spin zitterbewegung
tunneling with generalized-uncertainty-principle corrections. We derive the
particle-spin-dependent Hawking temperature, construct the corresponding
channel-dependent effective entropy through the first law, and formulate the
associated off-shell free energy and thermodynamic vector field. As reference
sectors, we also analyze
the Bekenstein-Hawking and exponentially corrected entropies. We find a clear
separation between local and global thermodynamic topology. For the
$\gamma=-1$ branch, a single positively wound defect persists, yielding
$W=+1$, whereas for $\gamma=+1$ the defects occur in oppositely wound pairs
and the global charge remains $W=0$. The particle spin nevertheless shifts
the defect positions and the critical configurations at which the
$\gamma=+1$ pair merges and annihilates. The fold separating the two-defect
and defect-free regions is controlled jointly by the particle spin, the GUP
parameter, and the cosmological constant. Thus, the quantum properties of the
emitted particle do not alter the global topological class, but they
reorganize the local defect and bifurcation structure within it. Our results
establish a direct connection between quantum-corrected Hawking radiation and the
corresponding effective off-shell thermodynamic topology, showing that its local
defect structure can retain information about the quantum channel through which the
black hole radiates.
\end{abstract}


\maketitle

\section{Introduction}
\label{isec1}

Black-hole thermodynamics provides a direct framework in which gravity, quantum
theory, and statistical mechanics meet. Since Hawking related black-hole
radiation to the surface gravity~\cite{Hawking:1975vcx} and Bekenstein
associated entropy with the horizon area~\cite{Bekenstein:1973ur}, black-hole
thermodynamics has become an important probe of the microscopic structure of
gravity~\cite{wald2001thermodynamics}. In the tunneling picture, Hawking
radiation is described as a quantum process across the horizon, with the
emission probability determined by the imaginary part of the particle action
~\cite{Parikh:1999mf}, allowing the properties of the emitted quantum to
enter explicitly into the radiation process.

A complementary perspective is provided by black-hole thermodynamic topology.
In the construction of Wei, Liu, and Mann, equilibrium configurations are
identified with defects of a thermodynamic vector field obtained from a
generalized off-shell free energy~\cite{WOS:001043962200006}. Each defect
carries a winding number and the sum defines a global topological charge $W$.
The framework has since been refined and extended in several directions
~\cite{Wei:2021vdx,Fang:2022rsb,Zhang:2023uay,Wei:2024gfz,Wu:2024asq,
Wei:2026upy,Yerra:2022coh,Ahmed:2022kyv,Hazarika:2024imk}, and applied to a
broad range of black-hole geometries and thermodynamic settings, including
Lovelock and Gauss-Bonnet gravity~\cite{WOS:000954813100003,
Tsilioukas:2023tdw,WOS:000989369500001,Anagnostopoulos:2025tax,
Petronikolou:2025mlm,Tsilioukas:2026gvy}, charged and dilatonic solutions
~\cite{WOS:001013054900009,WOS:001362395600002,Silva:2026mft},
Euler-Heisenberg AdS black holes~\cite{WOS:001210112800001}, and further
extensions~\cite{WOS:000964332800006,WOS:001643162100016,
WOS:001427635500001,WOS:001629426400006,
Wu:2022whe,Wu:2023sue,Gogoi:2023xzy,Zhu:2024jhw,Wu:2024rmv,
Chen:2024sow,Zhu:2024zcl,Wu:2023meo,Yerra:2022alz,Yerra:2022eov,
Sadeghi:2023aii,Yasir:2025hne}. Related work has examined modified
entropies and quantum-gravity-inspired corrections~\cite{Saridakis:2026vxo,
Anand:2025cer,Ladghami:2025rkc,Tsilioukas:2024seh,
Tello-Ortiz:2026aqa,BahrozBrzo:2026bbc,BahrozBrzo:2026gty,Usman:2026huz,
BahrozBrzo:2025pgt,Sekhmani:2025zwc,Anand:2026bza,
Ma:2026oqh}, as well as modified-gravity backgrounds
~\cite{Yin:2025rhr,Sadeghi:2026jic}.
The off-shell viewpoint is also closely related to generalized free-energy
landscapes~\cite{Carlip:1993sa,Li:2022oup}.

At the same time, quantum-gravitational effects are expected to modify Hawking
radiation. The generalized uncertainty principle introduces a minimal length
and modifies the tunneling probability and Hawking temperature
~\cite{tawfik2015review,Scardigli:2014qka,Das:2009hs,
das2011planck}, with numerous applications to black-hole backgrounds
~\cite{Li:2016mwq,Anacleto:2015mma,Casadio:2020rsj,Casadio:2020rsj,
Petruzziello:2020wkd,Singh:2023vol}. The emitted particles also carry mass,
angular momentum, and spin. Their spin content can be described through the
zitterbewegung formalism~\cite{Barut:1980mv,Barut:1985xp,Barut:1989aw},
which has been used in black-hole tunneling calculations
~\cite{Tekincay:2021gce}.
Tunneling has also been studied for fermionic, gravitino, vector, and tensor
modes~\cite{Kerner:2007rr,Kerner:2008qv,Yale:2008kx,Majhi:2009uk,
Siahaan:2015vwa,Erbin:2017zwo}, while recent analyses combining
zitterbewegung dynamics with GUP corrections have found explicitly
spin-dependent modifications of black-hole radiation and thermodynamics
~\cite{Sucu:2026njf}.

These two lines of investigation have nevertheless remained largely
disconnected. Thermodynamic topology is usually determined from the black-hole
geometry, entropy, and ensemble, whereas particle-dependent tunneling modifies
the Hawking temperature through the quantum numbers of the emitted particle.
This raises the central questions of the present work, namely whether the
information carried by
the emitted particle can propagate into the thermodynamic vector field and
reorganize its defect structure, and whether  the particle spin can shift
the
defects and their critical mergers, or even alter the global topological class.
Answering this question tests whether thermodynamic topology is fixed entirely
by the background geometry or can also retain information about the quantum
channel through which the black hole radiates.

We address this question for a static and spherically symmetric massive-gravity
AdS black hole. Massive-gravity black holes and their thermodynamic phase
structure have been studied extensively~\cite{Bebronne:2009mz,Fernando:2016sps,
Wu:2020fij,Zeng:2015tfj}, and their thermodynamic topology has been analyzed in
several settings~\cite{Chen:2023ddv,Zhang:2023tlq,Malik:2023nso,Chen:2024atr,
Mohamed:2024plq,Aslam:2025ren,Chen:2026ndo}. The geometry considered here
contains a charge-like term governed by an exponent $\lambda$ and a discrete
parameter $\gamma$. For $\lambda=2$, the $\gamma=-1$ branch reduces to the
Reissner-Nordstr\"om-AdS form, whereas the $\gamma=+1$ branch exhibits
Schwarzschild-AdS-like behavior, providing a useful setting in which two
different thermodynamic branches can be compared within the same framework. Then, we
combine the arbitrary-spin zitterbewegung formalism with a GUP deformation
to derive the particle-spin-dependent Hawking temperature. We then construct
the associated effective entropy from the first law and formulate a consistent
off-shell free energy with independent inverse temperature $\tau$. As reference
sectors, we also analyze the Bekenstein-Hawking entropy and the exponentially
corrected entropy of Ref.~\cite{chatterjee2020exponential}.

Our results reveal a clear distinction between local and global thermodynamic
topology. The particle spin and the GUP correction shift the defect positions
and the critical configurations at which oppositely wound defects merge or
annihilate, while the global topological charge remains fixed by the asymptotic
structure of the two branches. Thus, the global degree is protected, but the
organization of defects within a given topological class becomes particle-dependent.
This establishes a direct connection between quantum-corrected Hawking radiation and
the corresponding effective off-shell thermodynamic topology.

The manuscript is organized as follows. In Sec.~\ref{isec2} we introduce the
massive-gravity AdS background and derive the GUP-corrected
particle-spin-dependent Hawking temperature. In Sec.~\ref{isec3} we present the
thermodynamic-topology framework and the reference sectors. In
Sec.~\ref{isec4} we analyze the particle-spin-dependent thermodynamic field,
its defects, and its critical behavior. Finally, Sec.~\ref{Conclusions}
contains the conclusions.

\section{Massive-gravity AdS black hole and spin-dependent Hawking
radiation}
\label{isec2}

In this section we establish the gravitational and quantum framework that will
be used in the thermodynamic-topology analysis. We first introduce the
massive-gravity AdS black-hole background and its basic thermodynamic
quantities. We then study the tunneling of arbitrary-spin particles in the
presence of generalized-uncertainty-principle corrections and obtain the
corresponding particle-spin-dependent Hawking temperature.

\subsection{Massive-gravity AdS background}
\label{isec2A}

We begin by briefly reviewing the massive-gravity AdS black-hole solution that
will be used throughout the analysis. We consider a static and spherically
symmetric geometry described by the line element
\begin{equation}
ds^{2}=-f(r)\,dt^{2}+\frac{dr^{2}}{f(r)}
+r^{2}\left(d\theta^{2}+\sin^{2}\theta\,d\phi^{2}\right),
\label{eqmetric}
\end{equation}
where the metric function is
\begin{equation}
f(r)=1-\frac{2M}{r}
-\gamma\frac{Q^{2}}{r^{\lambda}}
-\frac{\Lambda r^{2}}{3}.
\label{eqmetricfunction}
\end{equation}
Here, $M$ and $Q$ denote the mass and charge parameters, respectively, while
$\lambda$ is an integration constant associated with the massive-gravity
sector. The effective cosmological constant is given by
\begin{equation}
\Lambda=2m_{g}^{2}\left(1-b^{6}\right),
\end{equation}
where $m_g$ denotes the graviton mass and $b$ is a model parameter. In the
following we restrict ourselves to $\Lambda<0$, corresponding to
asymptotically AdS geometries.

The discrete parameter $\gamma=\pm1$ distinguishes two branches of the
solution with different small-radius behavior. For $\gamma=+1$ the
charge-like contribution enters with the same sign as the mass term and the
geometry exhibits a Schwarzschild-AdS-like behavior. For $\gamma=-1$ the
charge-like term changes sign and the geometry becomes
Reissner-Nordstr\"om-AdS-like. In particular, for $\lambda=2$ its radial
dependence coincides with that of the standard Reissner-Nordstr\"om-AdS
solution. This choice therefore provides a simple reference case while
retaining the two distinct massive-gravity branches.

The event-horizon radius $r_h$ is determined by the condition $f(r_h)=0$.
Solving this relation for the mass parameter yields
\begin{equation}
M(r_h)=\frac{r_h}{2}
-\frac{\gamma Q^{2}}{2r_h^{\lambda-1}}
-\frac{\Lambda r_h^{3}}{6}.
\label{eqmass}
\end{equation}
The Hawking temperature follows from the surface gravity as
\begin{equation}
T_H=\frac{f'(r_h)}{4\pi},
\label{eqTHawking}
\end{equation}
while the corresponding Bekenstein-Hawking entropy is
\begin{equation}
S_{\rm BH}=\pi r_h^{2}.
\label{eqentropy}
\end{equation}

In the following, for the numerical analysis we adopt
$Q=1$, $\lambda=2$, and $\Lambda=-0.1$. The choice $\lambda=2$ allows a
direct comparison with the Reissner-Nordstr\"om-AdS-like branch discussed
above and will also simplify the analysis of the critical configurations. We
additionally introduce the dimensionless horizon radius
\begin{equation}
x\equiv\frac{r_h}{r_0},
\label{eqx}
\end{equation}
where $r_0$ is a reference length scale that will be set to unity in the
numerical calculations.

 \subsection{GUP-deformed zitterbewegung tunneling}
\label{isec2B}

We now investigate the tunneling of particles with arbitrary spin across the
black-hole horizon in the presence of quantum-gravitational corrections. The
spin dependence is incorporated through the zitterbewegung formalism, while
minimal-length effects are introduced through the generalized uncertainty
principle. Following Ref.~\cite{Tekincay:2021gce}, we take as our
starting point the generalized wave equation for spin-$s$ zitterbewegung modes
in curved spacetime
\begin{equation}
\begin{aligned}
-i\,\beta^{0}_{(s)}(x)\,\partial_{0}\tilde{\Psi}_{(s)}
&=
\Big[
i\,\beta^{n}_{(s)}(x)\partial_{n}
+i\,\beta^{\mu}_{(s)}(x)\Gamma_{\mu}^{(s)}
-\frac{m_s}{\hbar}
\Big]
\\
&\times
\left[
1+\beta_{\rm GUP}
\left(
\hbar^{2}\partial_{n}\partial^{n}
-m_s^{2}
\right)
\right]
\tilde{\Psi}_{(s)},
\end{aligned}
\label{eqzitter}
\end{equation}
where $\beta^\mu_{(s)}$ are the generalized spin matrices,
$\Gamma_\mu^{(s)}$ is the corresponding spin connection, $m_s$ denotes the
mass of the emitted spin-$s$ particle, and $\beta_{\rm GUP}$ denotes the GUP
deformation parameter associated with the minimal length.
 We employ
Eq.~\eqref{eqzitter} as the arbitrary-spin extension
of the zitterbewegung tunneling equation and focus on its semiclassical
near-horizon consequences.

In order to extract the tunneling rate, we adopt the WKB ansatz
\begin{equation}
\tilde{\Psi}_{(s)}
=
\exp\!\left(\frac{i}{\hbar}I_{(s)}\right)\psi_{(s)},
\label{eqwkb}
\end{equation}
where the action is decomposed into temporal, radial, and angular parts as
\begin{equation}
I_{(s)}
=
-\epsilon_{(s)}t
+W_{(s)}(r)
+\mathcal{Y}_{(s)}(\theta,\phi).
\label{eqaction}
\end{equation}
Substituting Eq.~\eqref{eqwkb} into Eq.~\eqref{eqzitter} and retaining the
leading-order terms in $\hbar$, we obtain the deformed Hamilton-Jacobi
equation. Its radial part can be written as
{\small{
\begin{equation}
f^{-1/2}\beta^{0}_{(s)}\epsilon_{(s)}\psi_{(s)}
=
\left[
\sqrt{f}\,\beta^{1}_{(s)}W'(r)+m_s
\right]
\left[
1+\beta_{\rm GUP}\tilde{X}
\right]
\psi_{(s)},
\label{eqradial}
\end{equation}}}
where
\begin{equation}
\tilde{X}
=
f(r)\left(W'(r)\right)^2-m_s^2.
\label{eqXtilde}
\end{equation}

Equation~\eqref{eqradial} is a matrix equation in the spin-$s$
zitterbewegung representation. Following the determinant reduction of the
multi-spin system developed in Refs.~\cite{Tekincay:2021gce,Sucu:2026njf}, one
obtains a scalar radial Hamilton-Jacobi condition for each spin sector.
Expanding this condition to first order in $\beta_{\rm GUP}$ and separating
the angular contribution through the constant $j$, the radial momentum can
be written in the form
{\small{
\begin{equation}
W'(r)
=
\pm
\frac{\sqrt{\epsilon_{(s)}^{2}-f(r)\mathcal{J}}}{f(r)}
\left[
1+\beta_{\rm GUP}\Upsilon_{(s)}(r)
+\mathcal{O}\!\left(\beta_{\rm GUP}^{2}\right)
\right],
\label{eqWprime}
\end{equation}}}
where
\begin{equation}
\mathcal{J}
\simeq
m_s^2+\frac{j^2}{r^2}.
\label{eqJdef}
\end{equation}
Here $j$ denotes the angular separation parameter associated with the angular
part of the semiclassical action.
The spin dependence originates from the determinant reduction of the generalized
matrix Hamilton-Jacobi equation. In the zitterbewegung representation adopted in
\cite{Tekincay:2021gce,Sucu:2026njf}, this reduction generates the characteristic
spin factor
\begin{equation}
m_s^2
\left[
1+\frac{2}{(4s)^2}
\right].
\label{eqspinfactor}
\end{equation}
Thus, this factor is a consequence of the particular arbitrary-spin
zitterbewegung construction employed here, and should not be interpreted as
a universal spin correction independent of the underlying generalized wave
equation.

For the tunneling calculation only the near-horizon value of the first-order
correction contributes to the residue. For the separated modes considered
here it is
\begin{equation}
\Upsilon_{(s)}(r_h)
=
m_s^2
\left[
1+\frac{2}{(4s)^2}
\right]
-\frac{j^2}{2r_h^2}.
\label{eqUpshorizon}
\end{equation}
Accordingly, the quantity entering the corrected temperature is fully
determined by the particle mass, spin, angular separation parameter, and
horizon radius.

The imaginary part of the action arises from the pole of the radial function
at the event horizon. Expanding the metric function around $r=r_h$ gives
\begin{equation}
f(r)\simeq f'(r_h)(r-r_h).
\label{eqfnear}
\end{equation}
Substituting this expansion into Eq.~\eqref{eqWprime} and evaluating the
outgoing contour around the simple pole yields
\begin{equation}
\mathrm{Im}\,W_{+}
=
\frac{\pi\epsilon_{(s)}}{f'(r_h)}
\left[
1+\beta_{\rm GUP}\Upsilon_{(s)}(r_h)
\right]
+\mathcal{O}\!\left(\beta_{\rm GUP}^{2}\right).
\label{eqImW}
\end{equation}
The tunneling probability is therefore
\begin{equation}
\Gamma
\sim
\exp\!\left(-4\,\mathrm{Im}\,W_{+}\right).
\label{eqGamma}
\end{equation}
Comparing Eq.~\eqref{eqGamma} with the corresponding Boltzmann factor yields
the GUP-corrected temperature
\begin{equation}
T_Z
=
\frac{f'(r_h)}{4\pi}
\frac{1}{1+\beta_{\rm GUP}\Upsilon_{(s)}(r_h)}.
\label{eqTZ}
\end{equation}
Moreover, using the standard Hawking temperature of Eq.~\eqref{eqTHawking},
this can be written to first order in $\beta_{\rm GUP}$ as
\begin{equation}
T_Z
=
T_H
\left[
1-\beta_{\rm GUP}\Upsilon_{(s)}(r_h)
+\mathcal{O}\!\left(\beta_{\rm GUP}^{2}\right)
\right].
\label{eqTZexpand}
\end{equation}
Hence, the standard Hawking temperature is recovered in the limit
$\beta_{\rm GUP}\rightarrow0$, while for nonvanishing $\beta_{\rm GUP}$ the
tunneling process acquires corrections that retain information about the
emitted particle. In the following subsection we introduce
a compact notation for $\Upsilon_{(s)}(r_h)$ and use the corresponding
particle-spin-dependent temperature in the thermodynamic-topology analysis.

\subsection{Particle-spin-dependent Hawking temperature}
\label{isec2C}

We now make explicit the dependence of the GUP-corrected Hawking temperature
on the quantum properties of the emitted particle. For convenience, we denote
the horizon correction appearing in Eq.~\eqref{eqUpshorizon} by
\begin{equation}
Y_{(s)}
\equiv
\Upsilon_{(s)}(r_h)
=
m_s^{2}
\left[
1+\frac{2}{(4s)^2}
\right]
-\frac{j^{2}}{2r_h^{2}},
\label{eqYs}
\end{equation}
where $m_s$ and $s$ denote the mass and spin of the emitted particle,
respectively, while $j$ is the angular separation parameter introduced in
subsection \ref{isec2B}. Thus, the first term in Eq.~\eqref{eqYs} carries the
explicit spin dependence inherited from the GUP-deformed zitterbewegung
formalism, while the second term arises from the angular separation sector.

Hence, to first order in the GUP parameter, the corrected
temperature becomes
\begin{equation}
T_{ZB}
=
T_H
\left[
1-\beta_{\rm GUP}Y_{(s)}
\right].
\label{eqTZcompact}
\end{equation}
Using the Hawking temperature of the massive-gravity AdS black hole, this can
equivalently be written as
\begin{equation}
T_{ZB}
=
\frac{1}{4\pi r_h}
\left[
1
+\gamma(\lambda-1)\frac{Q^2}{r_h^\lambda}
-\Lambda r_h^2
\right]
\left[
1-\beta_{\rm GUP}Y_{(s)}
\right].
\label{eqTZexplicit}
\end{equation}
Thus, the quantum-gravitational correction depends both on the black-hole
geometry, through $Q$, $\lambda$, $\gamma$, and $\Lambda$, and on the
properties of the emitted particle, through $m_s$, $s$, and $j$. In
particular, the explicitly spin-dependent contribution in
Eq.~\eqref{eqYs} decreases with increasing $s$, and therefore different
particle-spin sectors experience different corrections to the standard
Hawking temperature.

In order to illustrate this dependence, we adopt the
representative values $\beta_{\rm GUP}=0.1$, $m_s=1$, and $j=1$,
together with the black-hole parameters introduced
in subsection \ref{isec2A}. The value $\beta_{\rm GUP}=0.1$ is used as a
convenient benchmark that keeps the first-order correction
perturbative and allows its effect to be displayed clearly.
Our conclusions below will also be examined under variations
of $\beta_{\rm GUP}$.

The common values $m_s=1$ and $j=1$ are kept fixed across all
spin sectors in order to isolate the dependence on the particle
spin $s$. Thus, the comparison below should be understood as a
controlled fixed-mass, fixed-angular-mode comparison, rather
than as a comparison among realistic particle species with
different masses and angular separation parameters. For the
benchmark region relevant to the thermodynamic defects, the
quantity $|\beta_{\rm GUP}Y_{(s)}|$ remains at the level of
approximately $0.15$ or below, consistently with the first-order
perturbative character of the analysis.

Additionally, Table~\ref{tabtemperature} presents the values of $Y_{(s)}$ and the
corresponding ratio $T_{\rm ZB}/T_H$ for the four particle-spin
sectors $s=1/2,1,3/2,$ and $2$, evaluated at $r_h=2r_0$ and in
the large-horizon limit.

\begin{table}[ht!]
\centering
\begin{tabular}{ccccc}
\hline\hline
$s$
&
$Y_{(s)}|_{2r_0}$
&
$\dfrac{T_{ZB}}{T_H}\Big|_{2r_0}$
&
$Y_{(s)}|_{\infty}$
&
$\dfrac{T_{ZB}}{T_H}\Big|_{\infty}$
\\
\hline
$\tfrac12$ & $1.3750$ & $0.8625$ & $1.5000$ & $0.8500$ \\
$1$        & $1.0000$ & $0.9000$ & $1.1250$ & $0.8875$ \\
$\tfrac32$ & $0.9306$ & $0.9069$ & $1.0556$ & $0.8944$ \\
$2$        & $0.9062$ & $0.9094$ & $1.0312$ & $0.8969$ \\
\hline\hline
\end{tabular}
\caption{Spin-dependent GUP factor $Y_{(s)}$ and the corresponding ratio
$T_{ZB}/T_H$ for different particle-spin sectors. The results are shown at
$r_h=2r_0$ and in the limit $r_h\rightarrow\infty$, for
$\beta_{\rm GUP}=0.1$, $m_s=1$, and $j=1$.}
\label{tabtemperature}
\end{table}

\begin{figure}[ht!]
  \centering
  \includegraphics[width=\columnwidth]{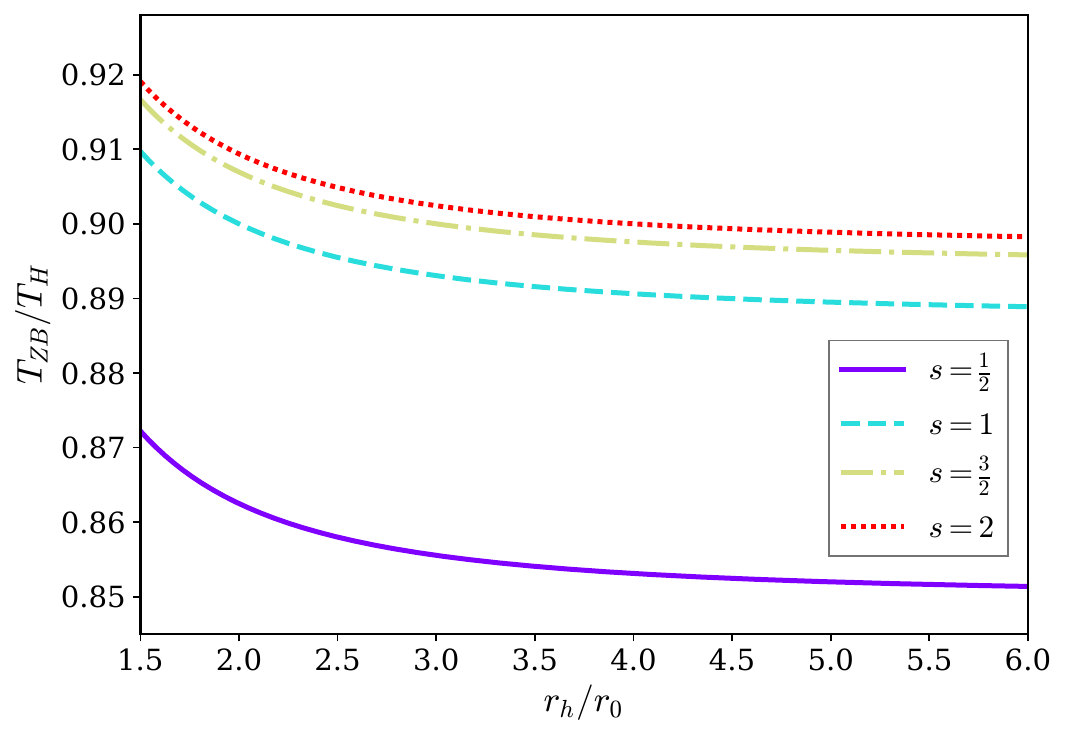}
  \caption{{\it{Ratio $T_{ZB}/T_H$ from Eq.~\eqref{eqTZcompact} as a function of
$r_h/r_0$ for the spin sectors $s=\tfrac12,1,\tfrac32,2$. The radial dependence
arises from the angular-momentum contribution proportional to $j^2/r_h^2$, while the
relative separation between the spin sectors is determined by the spin-dependent term
in $Y_{(s)}$. Parameters as in Table~\ref{tabtemperature}.}}}
  \label{figtemperature}
\end{figure}

As can be seen from Table~\ref{tabtemperature}, at $r_h=2r_0$ the ratio
$T_{ZB}/T_H$ increases from $0.8625$ for $s=1/2$ to $0.9094$ for $s=2$.
This hierarchy follows directly from the decrease of the term
$m_s^2[1+2/(4s)^2]$ with increasing particle spin. The same ordering is
maintained in the large-horizon limit, where the angular contribution
proportional to $j^2/r_h^2$ vanishes.

In order to provide the results in a more transparent way, in
Fig.~\ref{figtemperature} we present
the radial dependence. The term
$j^2/(2r_h^2)$ produces the same radius-dependent shift for all spin sectors,
whereas their relative separation is determined by the spin-dependent part
$m_s^2[1+2/(4s)^2]$. Hence, to the order considered here, the separation
between different spin curves does not depend on the horizon radius, although
their common radial evolution reflects the angular contribution in
Eq.~\eqref{eqYs}.

The resulting particle dependence of the Hawking temperature provides the
input for the thermodynamic-topology analysis below. In particular, the
corrected temperature will be used through the first law to construct the
corresponding effective entropy and generalized off-shell free energy. This
allows the particle-spin dependence to enter the local thermodynamic defect
structure in a manner consistent with the equilibrium condition.

\section{Thermodynamic topology and reference sectors}
\label{isec3}

Having established the particle-spin-dependent Hawking temperature, we now
introduce the thermodynamic-topology framework that will be used to determine
its consequences for the black-hole equilibrium structure. We first summarize
the off-shell construction and the associated topological invariants. Then
we apply it to the Bekenstein-Hawking and exponentially corrected entropy sectors,
which will provide the reference configurations for the particle-dependent
analysis of Sec. \ref{isec4}.

\subsection{Thermodynamic-topology framework}
\label{isec3A}

The thermodynamic-topology approach characterizes black-hole equilibrium
states as topological defects associated with the zeros of a vector field in
an auxiliary parameter space
~\cite{WOS:001043962200006,WOS:000964332800006}. The starting point is the
generalized off-shell free energy
\begin{equation}
\mathcal{F}
=
M-\frac{S}{\tau},
\label{eqFoffshell}
\end{equation}
where $M$ and $S$ denote the black-hole mass and entropy, respectively, while
$\tau$ is an independent inverse-temperature parameter. The on-shell
configuration is recovered when $\tau$ coincides with the inverse black-hole
temperature. Keeping $\tau$ independent allows the different thermodynamic
branches to be analyzed without imposing the equilibrium condition from the
outset.

Following the thermodynamic-topology construction, we introduce the
two-component vector field
\begin{equation}
\phi
=
\left(
\phi^{r},\phi^{\Theta}
\right),
\label{eqphidef}
\end{equation}
in the auxiliary space $(r_h,\Theta)$, with
\begin{equation}
\phi^{r}
=
\frac{\partial\mathcal{F}}{\partial r_h},
\qquad
\phi^{\Theta}
=
-\cot\Theta\,\csc\Theta.
\label{eqphicomp}
\end{equation}
The auxiliary angle lies in the interval $\Theta\in[0,\pi]$. Since
$\phi^{\Theta}=0$ at $\Theta=\pi/2$, the thermodynamic defects lie on this
line, while their radial positions are determined from
\begin{equation}
\phi^{r}=0.
\label{eqphirzero}
\end{equation}
Thus, the zeros of the vector field coincide with the equilibrium
configurations of the generalized free energy for fixed $\tau$.

In order to characterize the topology of these zeros, we introduce the
normalized vector field
\begin{equation}
n^{a}
=
\frac{\phi^{a}}{\|\phi\|},
\qquad
\|\phi\|
=
\sqrt{
(\phi^{r})^{2}
+
(\phi^{\Theta})^{2}
},
\qquad
a=1,2.
\label{eqnfield}
\end{equation}
The corresponding conserved topological current is
\begin{equation}
j^{\mu}
=
\frac{1}{2\pi}
\epsilon^{\mu\nu\rho}
\epsilon_{ab}
\partial_{\nu}n^{a}
\partial_{\rho}n^{b},
\qquad
\mu,\nu,\rho=0,1,2,
\label{eqcurrent}
\end{equation}
which satisfies
$
\partial_{\mu}j^{\mu}=0$.
Using Duan's $\phi$-mapping theory, the current can equivalently be written as
\begin{equation}
j^{\mu}
=
\delta^{2}(\phi)
J^{\mu}
\left(
\frac{\phi}{x}
\right),
\label{eqduan}
\end{equation}
where the Jacobian tensor is defined through
\begin{equation}
\epsilon^{ab}
J^{\mu}
\left(
\frac{\phi}{x}
\right)
=
\epsilon^{\mu\nu\rho}
\partial_{\nu}\phi^{a}
\partial_{\rho}\phi^{b}.
\label{eqjacobian}
\end{equation}
For $\mu=0$ this reduces to
\begin{equation}
J^{0}
\left(
\frac{\phi}{x}
\right)
=
\frac{\partial(\phi^{1},\phi^{2})}
{\partial(x^{1},x^{2})}.
\label{eqJ0}
\end{equation}

The topological current is therefore localized at the zeros of $\phi$.
Integrating its temporal component over a closed region $\Sigma$ in the
$(r_h,\Theta)$ plane gives the total topological charge
\begin{equation}
W
=
\int_{\Sigma}j^{0}d^{2}x
=
\sum_{i=1}^{N}\beta_i\eta_i
=
\sum_{i=1}^{N}\omega_i,
\label{eqWcharge}
\end{equation}
where $\beta_i$ is the Hopf index, $\eta_i$ is the Brouwer degree, and
$\omega_i=\beta_i\eta_i$ is the winding number of the $i$-th defect.

For the vector field of Eq.~\eqref{eqphicomp}, the Jacobian takes the form
\begin{equation}
J
=
\frac{\partial(\phi^{r},\phi^{\Theta})}
{\partial(r_h,\Theta)}
=
\begin{pmatrix}
\partial_{r_h}\phi^{r} & \partial_{\Theta}\phi^{r}
\\
\partial_{r_h}\phi^{\Theta} & \partial_{\Theta}\phi^{\Theta}
\end{pmatrix}.
\label{eqJacmatrix}
\end{equation}
Since $\phi^{r}$ depends only on $r_h$ and $\phi^{\Theta}$ only on $\Theta$,
the off-diagonal terms vanish. Hence,
\begin{equation}
\det J
=
\partial_{r_h}\phi^{r}
\partial_{\Theta}\phi^{\Theta},
\label{eqdetJ}
\end{equation}
with
\begin{equation}
\partial_{\Theta}\phi^{\Theta}
=
\frac{1+\cos^{2}\Theta}{\sin^{3}\Theta}.
\label{eqangularjac}
\end{equation}
At the defect line $\Theta=\pi/2$ one has
$\partial_{\Theta}\phi^{\Theta}=1$. Therefore, for a simple zero with
$\beta_i=1$, the winding number is determined directly by
\begin{equation}
\omega_i
=
{\rm sign}
\left[
\partial_{r_h}\phi^{r}
\right]_{r_h=r_i}.
\label{eqwindingsign}
\end{equation}
A zero with $\partial_{r_h}\phi^{r}>0$ corresponds to a local minimum of the
generalized free energy and carries $\omega=+1$, while a zero with
$\partial_{r_h}\phi^{r}<0$ corresponds to a local maximum and carries
$\omega=-1$. The two cases are associated with locally stable and unstable
thermodynamic branches, respectively.

The global character of the construction can also be expressed through the
degree of the normalized field. Writing
\begin{equation}
n
=
\left(
\cos\Omega,\sin\Omega
\right),
\end{equation}
the total charge can equivalently be obtained as
\begin{equation}
W
=
\frac{1}{2\pi}
\oint_{\partial\Sigma}
d\Omega
=
{\rm deg}
\left(
n|_{\partial\Sigma}
\right).
\label{eqdegree}
\end{equation}
Therefore, $W$ is invariant under smooth deformations of the thermodynamic
field as long as the boundary degree remains unchanged. The number of defects
may nevertheless change through the creation or annihilation of pairs with
opposite winding numbers. Such a merger occurs at a degenerate zero satisfying
\begin{equation}
\phi^{r}(r_c)=0,
\qquad
\partial_{r_h}\phi^{r}(r_c)=0.
\label{eqfoldgeneral}
\end{equation}
This fold condition will be used below to identify the critical configurations
associated with the particle-spin-dependent corrections.

\subsection{Bekenstein-Hawking entropy}
\label{isec3B}

We first consider the thermodynamic topology associated with the standard
Bekenstein-Hawking entropy, which will serve as the reference configuration
for the corrected sectors analyzed below. Using Eqs.~\eqref{eqmass}
and~\eqref{eqentropy}, the generalized off-shell free energy of
Eq.~\eqref{eqFoffshell} becomes
\begin{equation}
\mathcal{F}_{\rm BH}(r_h,\tau)
=
\frac{r_h}{2}
-\frac{\gamma Q^{2}}{2r_h^{\lambda-1}}
-\frac{\Lambda r_h^{3}}{6}
-\frac{\pi r_h^{2}}{\tau},
\label{eqFBH}
\end{equation}
and its radial thermodynamic field is
\begin{equation}
\phi^{r}_{\rm BH}
=
\frac{1}{2}
+\frac{\gamma Q^{2}(\lambda-1)}{2r_h^{\lambda}}
-\frac{\Lambda r_h^{2}}{2}
-\frac{2\pi r_h}{\tau}.
\label{eqphirBH}
\end{equation}
Since $\phi^\Theta=0$ fixes $\Theta=\pi/2$, the defect condition reads
\begin{equation}
1
+\frac{\gamma Q^{2}(\lambda-1)}{r_h^{\lambda}}
-\Lambda r_h^{2}
-\frac{4\pi r_h}{\tau}
=
0.
\label{eqzeroBH}
\end{equation}
Hence, solving for the inverse-temperature parameter gives
\begin{equation}
\tau(r_h)
=
\frac{4\pi r_h}
{1+\gamma Q^{2}(\lambda-1)r_h^{-\lambda}
-\Lambda r_h^{2}}.
\label{eqtauBH}
\end{equation}
Thus, comparison with Eq.~\eqref{eqTHawking} yields
\begin{equation}
\tau(r_h)=\frac{1}{T_H(r_h)},
\label{eqtauonBH}
\end{equation}
showing that the zeros of the off-shell thermodynamic field reproduce the
equilibrium black-hole configurations without imposing the Hawking-temperature
relation from the outset.

We now adopt the fiducial values $Q=1$, $\lambda=2$, and $\Lambda=-0.1$. The
two branches exhibit qualitatively different asymptotic behavior. For
$\gamma=-1$, which for $\lambda=2$ coincides with the
Reissner-Nordstr\"om-AdS form,
\begin{equation}
\phi^{r}_{\rm BH}\rightarrow-\infty
\qquad
{\rm for}
\qquad
r_h\rightarrow0^{+},
\end{equation}
whereas
\begin{equation}
\phi^{r}_{\rm BH}\rightarrow+\infty
\qquad
{\rm for}
\qquad
r_h\rightarrow\infty.
\end{equation}
Hence, the field possesses an odd number of zeros. For the parameter range
considered here the physical branch contains a single zero for every positive
$\tau$. In particular, for $\tau=18$ one finds
$
r_h\simeq5.094\,r_0$.
At this point
$
\partial_{r_h}\phi^{r}_{\rm BH}>0,
$
and therefore
$
\omega=+1$.
Since no additional defect is present, the total topological charge is
\begin{equation}
W=+1.
\label{eqWBHm1}
\end{equation}
This agrees with the standard thermodynamic-topology classification of the
Reissner-Nordstr\"om-AdS black hole.

For $\gamma=+1$, the radial field instead approaches positive values at both
ends of the physical radial domain,
\begin{equation}
\phi^{r}_{\rm BH}\rightarrow+\infty
\qquad
{\rm for}
\qquad
r_h\rightarrow0^{+},
\end{equation}
and
\begin{equation}
\phi^{r}_{\rm BH}\rightarrow+\infty
\qquad
{\rm for}
\qquad
r_h\rightarrow\infty.
\end{equation}
The defects can therefore occur only in pairs. At $\tau=18$,
Eq.~\eqref{eqzeroBH} admits two positive roots,
namely $
r_{h,1}\simeq2.631\,r_0$ and
$
r_{h,2}\simeq4.815\,r_0$, and the radial derivative is negative at the inner root
and positive at the outer
one. According to Eq.~\eqref{eqwindingsign}, their winding numbers are
$
\omega_1=-1$, and
$
\omega_2=+1$.
Thus, their contributions cancel and the total topological charge is
\begin{equation}
W=0.
\label{eqWBHp1}
\end{equation}

As $\tau$ increases, the two defects approach each other and eventually merge
at a degenerate configuration satisfying
\begin{equation}
\phi^{r}_{\rm BH}=0,
\qquad
\partial_{r_h}\phi^{r}_{\rm BH}=0.
\end{equation}
For the adopted parameter values this occurs at
\begin{equation}
r_c\simeq3.524\,r_0,
\qquad
\tau_c\simeq19.068.
\label{eqBHcritical}
\end{equation}
For $\tau<\tau_c$ two defects are present, at $\tau=\tau_c$ they merge, and
for $\tau>\tau_c$ no defect remains. Since the annihilating defects carry
opposite winding numbers, the total charge remains $W=0$ throughout the
transition.

\begin{figure}[t]
  \centering
  \includegraphics[width=\columnwidth]{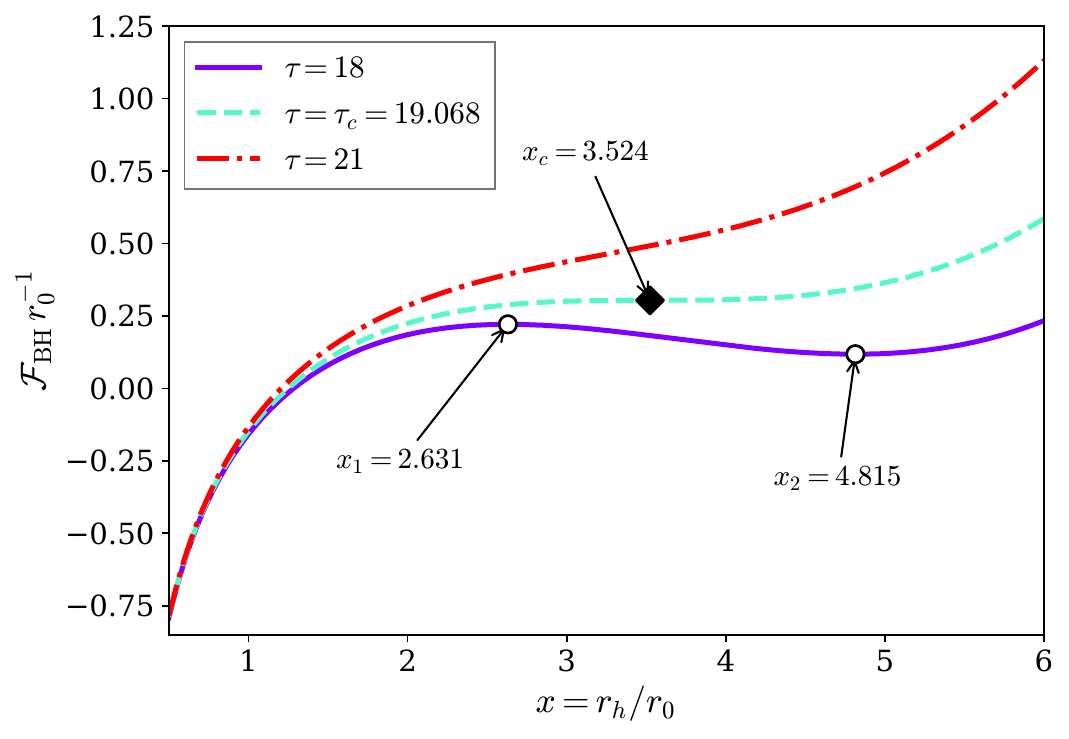}
 \caption{{\it{Off-shell free energy $\mathcal{F}=M-S/\tau$ of
Eq.~\eqref{eqFoffshell} as a function of $r_h/r_0$ for the Bekenstein-Hawking
entropy of the $\gamma=+1$ branch, for $\tau=18$, $\tau=\tau_c\simeq19.068$, and
$\tau=21$. For $\tau=18$ two stationary points are present at $r_h/r_0\simeq2.631$
and $4.815$. They merge at the critical configuration $r_c/r_0\simeq3.524$ for
$\tau=\tau_c$, while for $\tau=21$ no stationary point remains. Parameters as in
Sec.~\ref{isec2}.}}}
  \label{figfreeenergy}
\end{figure}

\begin{figure}[t]
  \centering
  \includegraphics[width=\columnwidth]{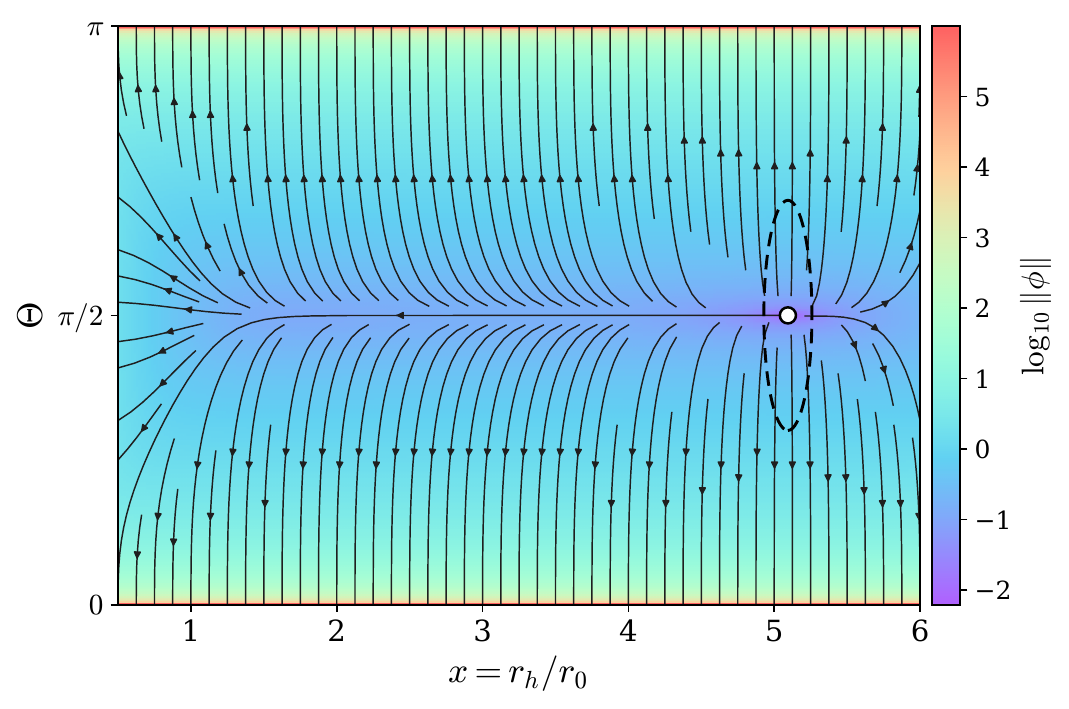}
 \caption{{\it{Normalized thermodynamic field $n=(n^r,n^\Theta)$ in the $(x,\Theta)$
plane for the $\gamma=-1$ Bekenstein-Hawking sector, with $x=r_h/r_0$ and $\tau=18$.
A single defect is located at $(x,\Theta)\simeq(5.094,\pi/2)$ and carries winding
number $\omega=+1$, yielding total topological charge $W=+1$. Parameters are $Q=1$,
$\lambda=2$, and $\Lambda=-0.1$.}}}
  \label{figstdgm1}
\end{figure}

\begin{figure}[t]
  \centering
  \includegraphics[width=\columnwidth]{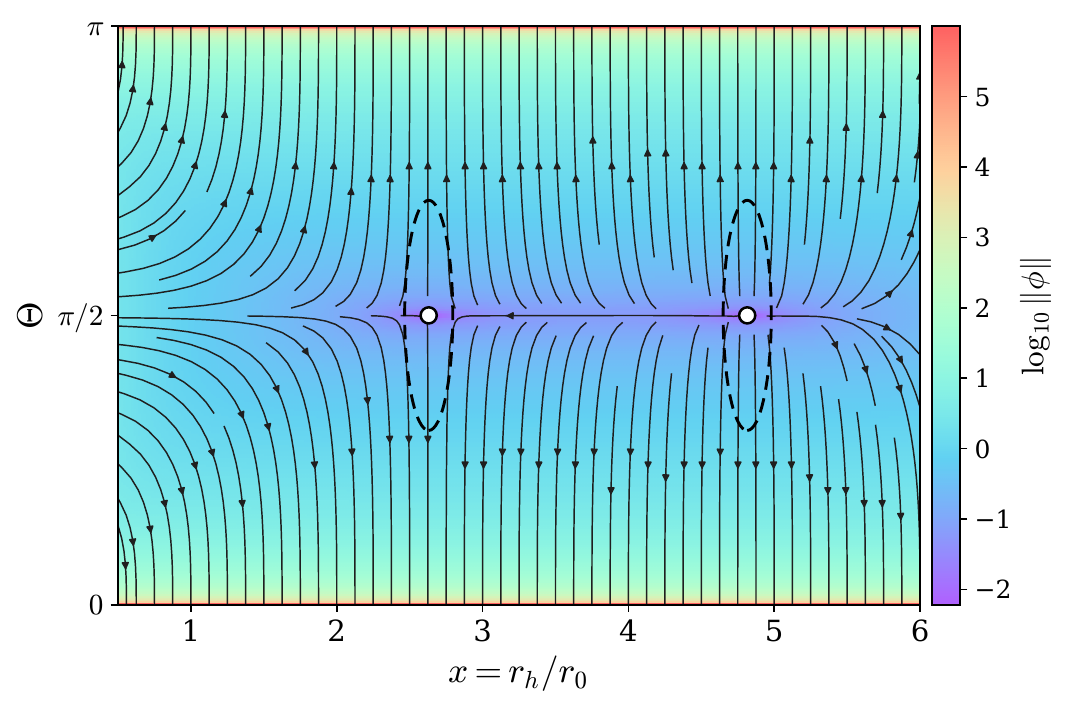}
 \caption{{\it{As in Fig.~\ref{figstdgm1} but for the $\gamma=+1$
Bekenstein-Hawking sector at $\tau=18$. Two defects are located at
$(x,\Theta)\simeq(2.631,\pi/2)$ and $(4.815,\pi/2)$, with winding numbers
$\omega_1=-1$ and $\omega_2=+1$, respectively. Their contributions cancel, yielding
total topological charge $W=0$.}}}
  \label{figstdgp1}
\end{figure}

Figure~\ref{figfreeenergy} illustrates this behavior through the off-shell free
energy. For the $\gamma=+1$ branch and $\tau<\tau_c$, the two stationary
points correspond to a local maximum and a local minimum with winding numbers
$-1$ and $+1$, respectively. They merge at the critical value of
Eq.~\eqref{eqBHcritical} and disappear for larger $\tau$. Furthermore, the
corresponding
thermodynamic vector fields are displayed in Figs.~\ref{figstdgm1}
and~\ref{figstdgp1}. As we can see, they show the single positively wound defect of
the
$\gamma=-1$ branch and the oppositely wound defect pair of the $\gamma=+1$
branch.

The Bekenstein-Hawking sector therefore provides the reference classification
\begin{equation}
W=
\begin{cases}
+1, & \gamma=-1,
\\
0, & \gamma=+1.
\end{cases}
\label{eqWBHsummary}
\end{equation}
In the following subsection we examine whether this global classification and
the associated local defect structure are modified when a quantum correction
is introduced directly through the black-hole entropy.

\subsection{Exponentially corrected entropy}
\label{isec3C}

We next examine whether the reference topological structure is modified when
quantum corrections are introduced directly through the black-hole entropy.
Following Ref.~\cite{chatterjee2020exponential}, we consider the exponentially
corrected entropy
\begin{equation}
S_{\rm exp}
=
S_{\rm BH}
+
e^{-S_{\rm BH}},
\label{eqSexp}
\end{equation}
which in the units adopted here becomes
\begin{equation}
S_{\rm exp}
=
\pi r_h^{2}
+
e^{-\pi r_h^{2}}.
\label{eqSexpr}
\end{equation}
The exponential contribution is relevant mainly at small horizon radii and
becomes rapidly suppressed as the horizon grows.

Using the mass function of Eq.~\eqref{eqmass}, the generalized off-shell free
energy takes the form
\begin{equation}
\mathcal{F}_{\rm exp}(r_h,\tau)
=
\frac{r_h}{2}
-\frac{\gamma Q^{2}}{2r_h^{\lambda-1}}
-\frac{\Lambda r_h^{3}}{6}
-\frac{1}{\tau}
\left(
\pi r_h^{2}
+
e^{-\pi r_h^{2}}
\right).
\label{eqFexp}
\end{equation}
The corresponding radial component of the thermodynamic field is
\begin{equation}
\phi^{r}_{\rm exp}
=
\frac{1}{2}
+
\frac{\gamma Q^{2}(\lambda-1)}{2r_h^{\lambda}}
-
\frac{\Lambda r_h^{2}}{2}
-
\frac{2\pi r_h}{\tau}
\left(
1-e^{-\pi r_h^{2}}
\right).
\label{eqphirexp}
\end{equation}
The angular component remains that of Eq.~\eqref{eqphicomp}, and therefore the
defects lie on $\Theta=\pi/2$ and are determined by the zeros of
Eq.~\eqref{eqphirexp}.

Solving the defect condition for the inverse-temperature parameter gives
\begin{equation}
\tau_{\rm exp}(r_h)
=
\frac{
4\pi r_h
\left(
1-e^{-\pi r_h^{2}}
\right)
}{
1+\gamma Q^{2}(\lambda-1)r_h^{-\lambda}
-\Lambda r_h^{2}
}.
\label{eqtauexp}
\end{equation}
Hence, comparison with Eq.~\eqref{eqtauBH} yields
\begin{equation}
\tau_{\rm exp}(r_h)
=
\tau_{\rm BH}(r_h)
\left(
1-e^{-\pi r_h^{2}}
\right).
\label{eqtauexprel}
\end{equation}
Thus, the exponential correction modifies the off-shell equilibrium curve
predominantly in the small-horizon regime, while the Bekenstein-Hawking
result is recovered for $r_h\rightarrow\infty$.

We now adopt the same parameter values as in the previous subsection. For the
$\gamma=-1$ branch, the radial field diverges negatively for
$r_h\rightarrow0^{+}$ and positively for $r_h\rightarrow\infty$. Hence, the
number of zeros remains odd. For the physical branch at $\tau=18$ one finds a
single defect at
$
r_h\simeq5.094\,r_0$.
At this point $\partial_{r_h}\phi^{r}_{\rm exp}>0$, and therefore
\begin{equation}
\omega=+1,
\qquad
W=+1.
\label{eqWexpm1}
\end{equation}

For the $\gamma=+1$ branch, the radial field is positive at both ends of the
physical radial domain. Consequently, the defects can occur only in pairs. At
$\tau=18$ the two positive roots are
$
r_{h,1}\simeq2.631\,r_0$, and
$
r_{h,2}\simeq4.815\,r_0$.
The inner root satisfies $\partial_{r_h}\phi^{r}_{\rm exp}<0$, whereas the
outer root satisfies $\partial_{r_h}\phi^{r}_{\rm exp}>0$. Their winding
numbers are therefore
\begin{equation}
\omega_1=-1,
\qquad
\omega_2=+1,
\end{equation}
and the total topological charge is
\begin{equation}
W=0.
\label{eqWexpp1}
\end{equation}

For the parameter values considered here, the exponential contribution is
already strongly suppressed at the radii of the thermodynamic defects.
Accordingly, their positions are indistinguishable, to the quoted numerical
precision, from those obtained in the Bekenstein-Hawking sector. Nevertheless,
Eq.~\eqref{eqtauexprel} shows explicitly that the off-shell thermodynamic
structure is modified at sufficiently small horizon radii.

The exponentially corrected sector therefore preserves the global
classification of the Bekenstein-Hawking reference case
\begin{equation}
W=
\begin{cases}
+1, & \gamma=-1,
\\
0, & \gamma=+1.
\end{cases}
\label{eqWexpsummary}
\end{equation}
We mention that we treat this correction as a reference quantum-entropy sector
distinct from
the GUP-induced particle-dependent correction that will be considered in Sec.
\ref{isec4} below. The latter
enters through the Hawking radiation and consequently provides a direct test of
how the quantum properties of the emitted particle affect the local
thermodynamic-topological structure.

\section{Particle-spin-dependent thermodynamic topology}
\label{isec4}

Having established the thermodynamic-topological structure of the reference
sectors, we now turn to the particle-spin-dependent case. We construct the
off-shell thermodynamic quantities associated with the GUP-corrected Hawking
temperature and determine the corresponding thermodynamic field. We then
analyze how the particle spin modifies the defect structure, the critical
configurations, and the resulting local and global topological properties.

\subsection{Spin-dependent free energy and thermodynamic field}
\label{isec4A}

We begin by constructing the thermodynamic description associated with the
particle-spin-dependent Hawking temperature derived in Sec.~\ref{isec2}. In order to
obtain the corresponding off-shell free energy consistently, we first
determine the effective entropy associated with the corrected temperature.    For
fixed
black-hole parameters, the first law can be written as
\begin{equation}
dM
=
T_{ZB}\,dS_{(s)}.
\label{eqfirstlawspin}
\end{equation}
Using Eq.~\eqref{eqTZcompact} together with the standard relation
\begin{equation}
dM
=
T_H\,dS_{\rm BH},
\end{equation}
we obtain, to first order in $\beta_{\rm GUP}$,
\begin{equation}
dS_{(s)}
=
\frac{dS_{\rm BH}}
{1-\beta_{\rm GUP}Y_{(s)}}
\simeq
\left[
1+\beta_{\rm GUP}Y_{(s)}
\right]
dS_{\rm BH}.
\label{eqdSspin}
\end{equation}
For convenience, we introduce
\begin{equation}
A_{(s)}
=
m_s^{2}
\left[
1+\frac{2}{(4s)^2}
\right],
\label{eqAsp}
\end{equation}
such that
\begin{equation}
Y_{(s)}
=
A_{(s)}
-\frac{j^{2}}{2r_h^{2}}.
\label{eqYspincompact}
\end{equation}
Since $S_{\rm BH}=\pi r_h^2$, integration of Eq.~\eqref{eqdSspin} yields
\begin{eqnarray}
&&
\!\!\!\!\!\!\!\!\!\!\!\!\!
S_{(s)}
=
\pi r_h^{2}
\left[
1+\beta_{\rm GUP}A_{(s)}
\right]
-
\pi\beta_{\rm GUP}j^{2}
\ln\left(\frac{r_h}{r_0}\right)\nonumber\\
&&
\ \
+
\mathcal{O}\left(\beta_{\rm GUP}^{2}\right),
\label{eqSspin}
\end{eqnarray}
where the irrelevant additive integration constant has been omitted and $r_0$ is the
reference scale introduced in subsection \ref{isec2A}. Here, $S_{(s)}$ should be
understood as a channel-dependent effective thermodynamic entropy associated with the
corresponding GUP-corrected tunneling sector, i.e. it is not interpreted as an
intrinsic microscopic entropy of the black hole or as a replacement for the
Bekenstein-Hawking entropy $S_{\rm BH}$. Accordingly, the off-shell free energy
introduced below is an effective thermodynamic potential for the specified emission
channel. Its role is to propagate the particle dependence of the corrected tunneling
temperature into the off-shell thermodynamic construction, while the underlying
black-hole geometry and the Bekenstein-Hawking entropy remain unchanged. As expected,
the standard Bekenstein-Hawking description is continuously recovered in the limit
$\beta_{\rm GUP}\to0$.

Therefore, the  generalized off-shell free energy for each particle-spin sector is
\begin{equation}
\mathcal{F}_{(s)}(r_h,\tau)
=
M(r_h)
-
\frac{S_{(s)}(r_h)}{\tau}.
\label{eqFspin}
\end{equation}
Using Eqs.~\eqref{eqmass} and~\eqref{eqSspin}, it becomes
\begin{align}
\mathcal{F}_{(s)}
=&
\frac{r_h}{2}
-
\frac{\gamma Q^{2}}{2r_h^{\lambda-1}}
-
\frac{\Lambda r_h^{3}}{6}
\nonumber
-
\frac{\pi r_h^{2}}{\tau}
\left[
1+\beta_{\rm GUP}A_{(s)}
\right]\nonumber
\\
&+
\frac{\pi\beta_{\rm GUP}j^{2}}{\tau}
\ln\left(\frac{r_h}{r_0}\right)
+
\mathcal{O}\left(\beta_{\rm GUP}^{2}\right).
\label{eqFspinexplicit}
\end{align}

Following the construction of subsection \ref{isec3A}, we introduce the
thermodynamic vector field
\begin{equation}
\phi_{(s)}
=
\left(
\phi^{r}_{(s)},
\phi^{\Theta}
\right),
\label{eqphispindef}
\end{equation}
with angular component
\begin{equation}
\phi^{\Theta}
=
-\cot\Theta\,\csc\Theta,
\end{equation}
and radial component
\begin{align}
\phi^{r}_{(s)}
=&
\frac{1}{2}
+
\frac{\gamma Q^{2}(\lambda-1)}
{2r_h^{\lambda}}
-
\frac{\Lambda r_h^{2}}{2}
\nonumber\\
&\!
-
\frac{2\pi r_h}{\tau}
\left[
1+\beta_{\rm GUP}A_{(s)}
-\frac{\beta_{\rm GUP}j^{2}}{2r_h^{2}}
\right]
+
\mathcal{O}\left(\beta_{\rm GUP}^{2}\right).
\label{eqphirspin}
\end{align}
Hence, the explicit spin dependence of the thermodynamic field enters through
the combination
\begin{equation}
\beta_{\rm GUP}A_{(s)}
=
\beta_{\rm GUP}m_s^{2}
\left(
1+\frac{2}{(4s)^2}
\right).
\label{eqspincombination}
\end{equation}

The defects are again located on $\Theta=\pi/2$, and their radial positions
follow from $\phi^{r}_{(s)}=0$. Solving this condition for the
inverse-temperature parameter gives
\begin{equation}
\tau_{(s)}(r_h)
=
\frac{
4\pi r_h
\left[
1+\beta_{\rm GUP}Y_{(s)}
\right]
}{
1+\gamma Q^{2}(\lambda-1)r_h^{-\lambda}
-\Lambda r_h^{2}
}
+
\mathcal{O}\left(\beta_{\rm GUP}^{2}\right).
\label{eqtausp}
\end{equation}
Using Eq.~\eqref{eqTZexplicit}, one immediately obtains
\begin{equation}
\tau_{(s)}(r_h)
=
\frac{1}{T_{ZB}(r_h)}
+
\mathcal{O}\left(\beta_{\rm GUP}^{2}\right).
\label{eqtauTspin}
\end{equation}
Thus, the zeros of the spin-dependent thermodynamic field coincide with the
equilibrium configurations determined by the GUP-corrected Hawking
temperature. This also provides a consistency check of the off-shell
construction.

Equation~\eqref{eqphirspin} constitutes the central quantity for the analysis
that follows. In the limit $\beta_{\rm GUP}\rightarrow0$ it reduces
continuously to the Bekenstein-Hawking field of Eq.~\eqref{eqphirBH}, while
for nonvanishing $\beta_{\rm GUP}$ different particle-spin sectors generate
different thermodynamic fields. Their defect positions and critical
configurations can therefore depend explicitly on the quantum properties of
the emitted particle, while the corresponding topological classification will
be determined in the following subsections.

\begin{figure*}[t]
  \centering
  \includegraphics[width=0.9\textwidth]{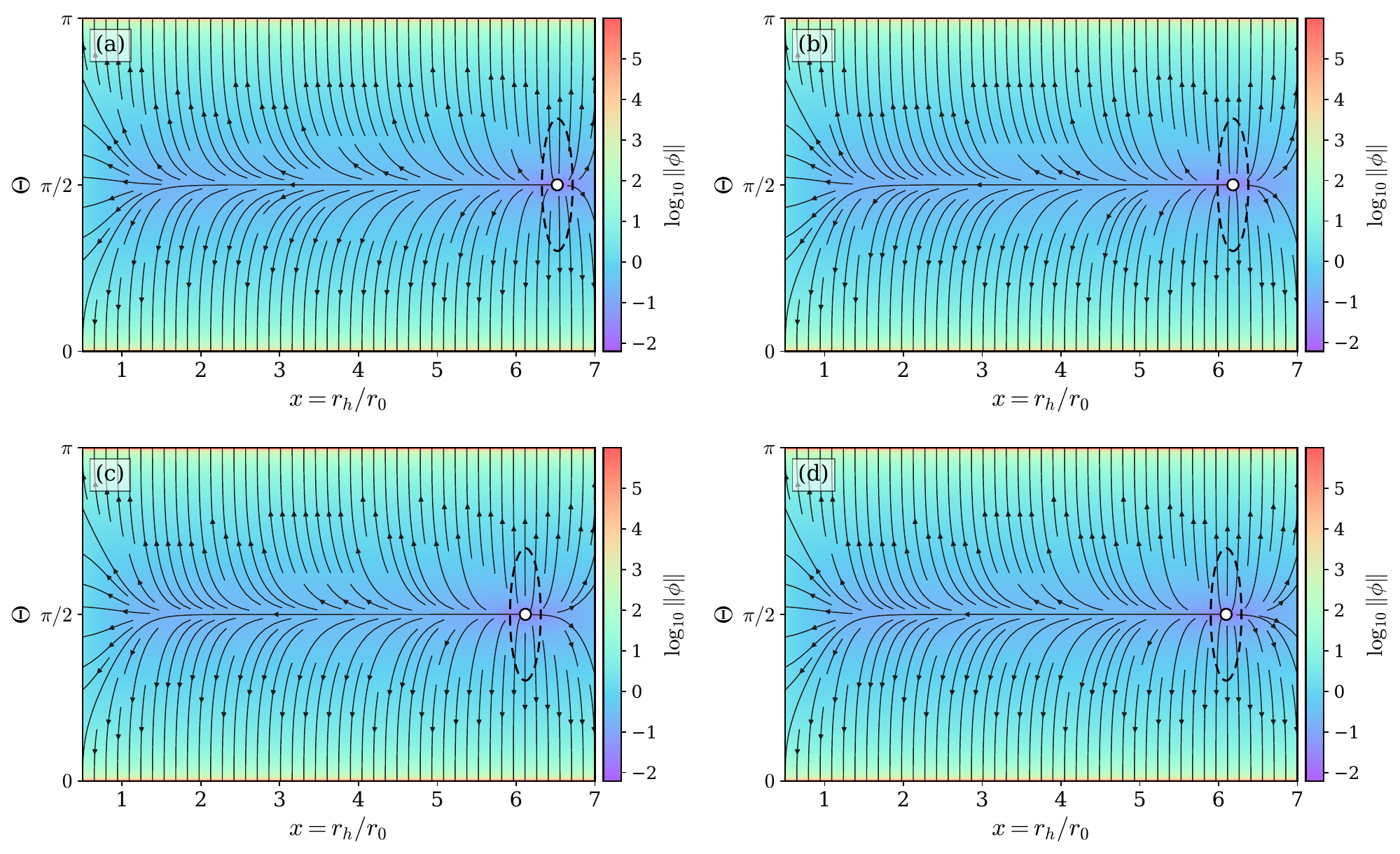}
  \caption{{\it{Normalized thermodynamic fields in the $(x,\Theta)$ plane for the
spin sectors $s=\tfrac12,1,\tfrac32,2$ shown in panels (a)-(d) for the $\gamma=-1$
branch at $\tau=18$, obtained from the zeros of Eq.~\eqref{eqphirspin}. Each spin
sector contains a single defect with winding number $\omega=+1$, while the defect
position shifts systematically with the particle spin. The total topological charge
remains $W=+1$ in all cases. Parameters are $Q=1$, $\lambda=2$, $\Lambda=-0.1$,
$\beta_{\rm GUP}=0.1$, $m_s=1$, $j=1$, and $r_0=1$.}}}
  \label{figspingm1}
\end{figure*}

\begin{figure*}[t]
  \centering
  \includegraphics[width=0.9\textwidth]{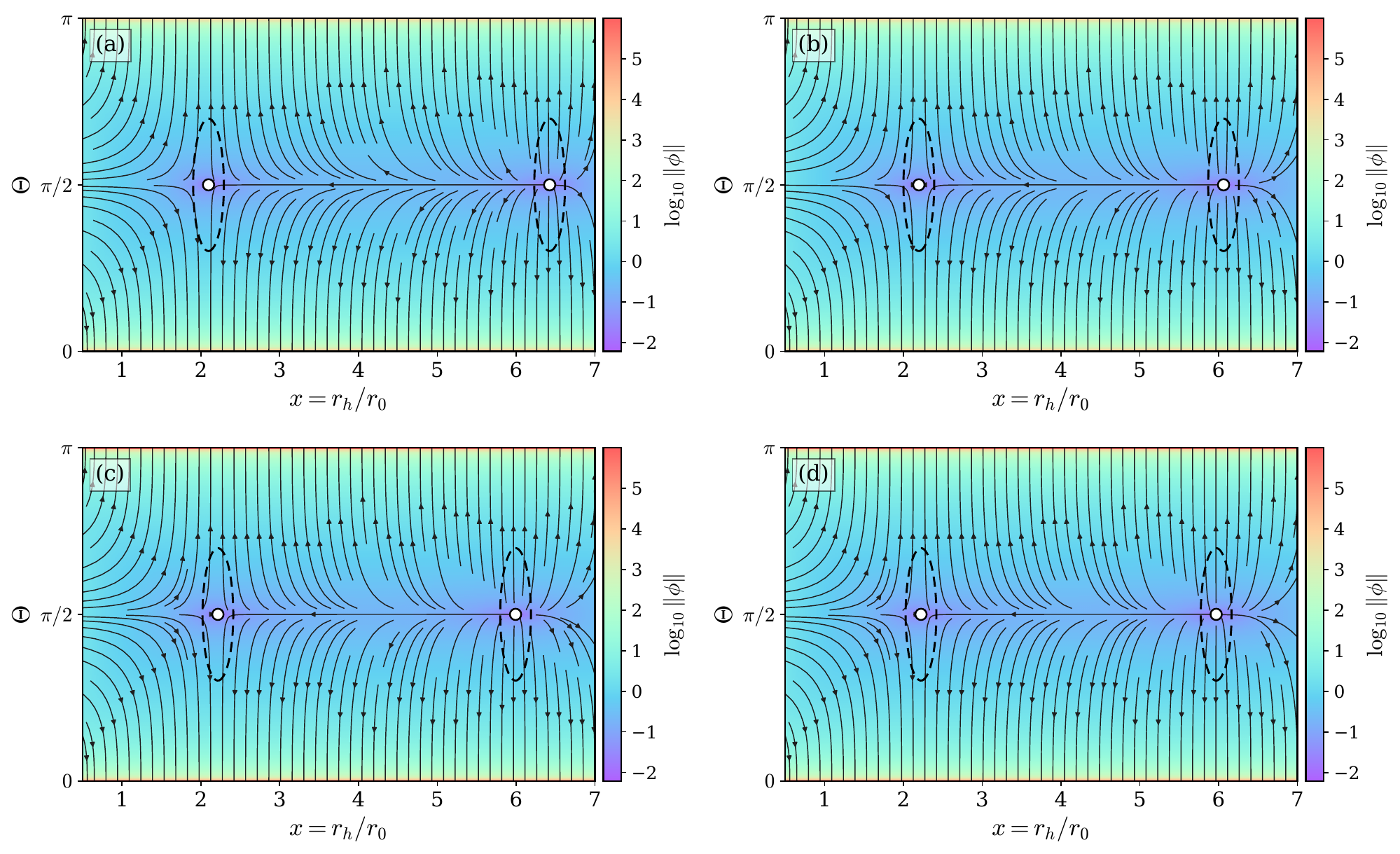}
 \caption{{\it{As in Fig.~\ref{figspingm1} but for the $\gamma=+1$ branch at
$\tau=18$. Each spin sector contains two thermodynamic defects, with the inner defect
carrying winding number $\omega=-1$ and the outer defect carrying $\omega=+1$. Their
positions depend on the particle spin, while their contributions cancel and the total
topological charge remains $W=0$ in all cases. Parameters as in
Fig.~\ref{figspingm1}.}}}
  \label{figspingp1}
\end{figure*}

\subsection{Spin-dependent defects and topological charge}
\label{isec4B}

Having obtained the spin-dependent thermodynamic field, we now investigate its
zeros and the associated winding numbers. We consider the representative
particle-spin values $s=1/2$, $1$, $3/2$, and $2$, while fixing the black-hole
and GUP parameters to $Q=1$, $\lambda=2$, $\Lambda=-0.1$,
$\beta_{\rm GUP}=0.1$, $m_s=1$, and $j=1$. In order to compare the different
spin sectors at the same thermodynamic state, we additionally set $\tau=18$.
The zeros are searched over the physically admissible radial domain, additionally
restricted by the perturbative requirement $|\beta_{\rm GUP}Y_{(s)}|\ll1$ associated
with the first-order GUP expansion. For the fiducial values $Q=1$, $\lambda=2$, and
$\Lambda=-0.1$, the $\gamma=-1$ branch possesses an extremal lower boundary
determined by $T_H=0$, namely $r_h/r_0\simeq0.957$. For the $\gamma=+1$ branch the
Hawking temperature remains positive, while imposing $M>0$ gives
$r_h/r_0\gtrsim0.984$. For $\beta_{\rm GUP}=0.1$ and the spin sectors considered
below, $|\beta_{\rm GUP}Y_{(s)}|$ remains smaller than approximately $0.15$
throughout the physically relevant region.

Before presenting the numerical results, it is nevertheless useful to examine the
formal limiting behavior of the truncated radial field.  For the adopted value
$\lambda=2$,
Eq.~\eqref{eqphirspin} gives, as $r_h\rightarrow0^{+}$,
\begin{equation}
\phi^{r}_{(s)}
\simeq
\frac{\gamma Q^{2}(\lambda-1)}
{2r_h^{\lambda}},
\label{eqspinlimitsmall}
\end{equation}
whereas for $r_h\rightarrow\infty$ the cosmological contribution dominates,
leading to
\begin{equation}
\phi^{r}_{(s)}
\simeq
-\frac{\Lambda r_h^{2}}{2}.
\label{eqspinlimitlarge}
\end{equation}
We stress that the literal $r_h\to0$ limit should not be interpreted as part of the
controlled first-order GUP regime, since $Y_{(s)}$ contains a term proportional to
$r_h^{-2}$. Equation (\ref{eqspinlimitsmall}) therefore indicates only the formal
leading behavior of the truncated field. For the actual classification we use the
lower boundary of the physically admissible perturbative domain. For the parameter
values considered here, direct evaluation at this boundary gives $\phi^r_{(s)}<0$ for
$\gamma=-1$ and $\phi^r_{(s)}>0$ for $\gamma=+1$, while Eq. (\ref{eqspinlimitlarge})
 gives $\phi^r_{(s)}>0$ at large radius. The resulting odd/even defect
classification is therefore unchanged.

Since $\Lambda<0$, the field approaches positive values at large radius for
both branches. At small radius its sign is determined by $\gamma$. Therefore,
for $\gamma=-1$ the field evolves from negative to positive values and must
contain an odd number of zeros, whereas for $\gamma=+1$ it is positive at both
boundaries and its zeros can occur only in pairs. The GUP and spin-dependent
terms are subleading in these limits and therefore do not alter the boundary
behavior of the field. Consequently, the global topological class remains the
same as in the reference Bekenstein-Hawking sector.

For the $\gamma=-1$ branch, we find one isolated defect for each of the four
spin sectors. Their positions are
\begin{equation}
\begin{aligned}
s=\frac12 \qquad & r_h\simeq6.523\,r_0,
\\
s=1 \qquad & r_h\simeq6.182\,r_0,
\\
s=\frac32 \qquad & r_h\simeq6.118\,r_0,
\\
s=2 \qquad & r_h\simeq6.095\,r_0.
\end{aligned}
\label{eqspinrootsm1}
\end{equation}
In all cases the derivative of the radial field at the zero is positive.
Hence, according to Eq.~\eqref{eqwindingsign},
\begin{equation}
\omega=+1,
\end{equation}
and consequently
\begin{equation}
W=+1
\qquad
{\rm for}
\qquad
\gamma=-1.
\label{eqWspinm1}
\end{equation}
Thus, increasing the particle spin shifts the thermodynamic defect towards
smaller horizon radii, while its winding number and local stability remain
unchanged.

The $\gamma=+1$ branch exhibits a different defect structure. For each spin
sector two defects are present, with positions
\begin{equation}
\begin{aligned}
s=\frac12 \qquad &
r_{h,1}\simeq2.097\,r_0,
\qquad
r_{h,2}\simeq6.426\,r_0,
\\
s=1 \qquad &
r_{h,1}\simeq2.196\,r_0,
\qquad
r_{h,2}\simeq6.063\,r_0,
\\
s=\frac32 \qquad &
r_{h,1}\simeq2.216\,r_0,
\qquad
r_{h,2}\simeq5.994\,r_0,
\\
s=2 \qquad &
r_{h,1}\simeq2.223\,r_0,
\qquad
r_{h,2}\simeq5.969\,r_0.
\end{aligned}
\label{eqspinrootsp1}
\end{equation}
The inner defect satisfies
$
\partial_{r_h}\phi^{r}_{(s)}<0$,
whereas the outer defect satisfies
$
\partial_{r_h}\phi^{r}_{(s)}>0$.
Their winding numbers are therefore
\begin{equation}
\omega_1=-1,
\qquad
\omega_2=+1,
\end{equation}
and the total charge is
\begin{equation}
W=\omega_1+\omega_2=0
\qquad
{\rm for}
\qquad
\gamma=+1.
\label{eqWspinp1}
\end{equation}

The corresponding normalized thermodynamic fields are displayed in
Figs.~\ref{figspingm1} and~\ref{figspingp1},
while the numerical results are summarized in Table~\ref{tabspindefects}. For
$\gamma=-1$, the single
positively wound defect moves continuously towards smaller $r_h$ as the
particle spin increases. For $\gamma=+1$, the two oppositely wound defects
also shift systematically, with the inner defect moving outward and the outer
one moving inward as $s$ increases. Thus, increasing the spin reduces the
separation between the two defects of this branch.

\begin{table}[ht!]
\centering
\begin{tabular}{ccccc}
\hline\hline
$s$
&
\multicolumn{2}{c}{$\gamma=-1$}
&
\multicolumn{2}{c}{$\gamma=+1$}
\\
&
$r_h/r_0$
&
$\omega$
&
$r_h/r_0$
&
$\omega$
\\
\hline
$\tfrac12$
&
$6.523$
&
$+1$
&
$2.097,\ 6.426$
&
$-1,\ +1$
\\
$1$
&
$6.182$
&
$+1$
&
$2.196,\ 6.063$
&
$-1,\ +1$
\\
$\tfrac32$
&
$6.118$
&
$+1$
&
$2.216,\ 5.994$
&
$-1,\ +1$
\\
$2$
&
$6.095$
&
$+1$
&
$2.223,\ 5.969$
&
$-1,\ +1$
\\
\hline\hline
\end{tabular}
\caption{Positions and winding numbers of the thermodynamic defects for the four
particle-spin sectors at $\tau=18$. The remaining parameters are $Q=1$,
$\lambda=2$, $\Lambda=-0.1$, $\beta_{\rm GUP}=0.1$, $m_s=1$, $j=1$, and
$r_0=1$.}
\label{tabspindefects}
\end{table}

These results show that the particle spin modifies the local
thermodynamic-topological structure while leaving the global charge of each
branch unchanged. In particular, the defect positions and therefore the
associated equilibrium configurations depend explicitly on the spin of the
emitted particle. The global classification remains
\begin{equation}
W=
\begin{cases}
+1, & \gamma=-1,
\\
0, & \gamma=+1,
\end{cases}
\label{eqWspinsummary}
\end{equation}
for all four spin sectors considered here. In the following subsection we
investigate the critical behavior underlying the displacement, merging, and
annihilation of these defects and its dependence on the particle and GUP
parameters.

\subsection{Topological transitions and parameter dependence}
\label{isec4C}

We now investigate how the particle spin and the GUP parameter affect the
merging and annihilation of thermodynamic defects. According to the general
discussion of subsection \ref{isec3A}, a pair of defects merges at a critical
configuration satisfying
\begin{equation}
\phi^{r}_{(s)}(r_c)=0,
\qquad
\left.
\frac{\partial\phi^{r}_{(s)}}{\partial r_h}
\right|_{r_h=r_c}
=0.
\label{eqspincritical}
\end{equation}
At such a point the Jacobian vanishes and two defects with opposite winding
numbers merge. Their annihilation changes the local defect structure while
preserving the total topological charge.

In order to analyze this behavior explicitly, we specialize to $\lambda=2$
and introduce
\begin{equation}
C_{(s)}
=
1+\beta_{\rm GUP}A_{(s)},
\label{eqCs}
\end{equation}
where $A_{(s)}$ is given in Eq.~\eqref{eqAsp}. The equilibrium relation of
Eq.~\eqref{eqtausp} then becomes
\begin{equation}
\tau_{(s)}(r_h)
=
\frac{
4\pi r_h
\left[
C_{(s)}
-\dfrac{\beta_{\rm GUP}j^2}{2r_h^2}
\right]
}{
1+\dfrac{\gamma Q^2}{r_h^2}
-\Lambda r_h^2
}.
\label{eqtauslambda2}
\end{equation}
The critical configurations correspond to extrema of this curve. Introducing
$u_c=r_c^2$, the condition $d\tau_{(s)}/dr_h=0$ yields
\begin{eqnarray}
&&
\!\!\!\!\!\!\!\!\!\!\!\!\!\!\!\!\!\!
2\Lambda C_{(s)}u_c^3
+
\left[
2C_{(s)}
-3\Lambda\beta_{\rm GUP}j^2
\right]
u_c^2
-
\gamma\beta_{\rm GUP}j^2Q^2
\nonumber\\
&& \  \, \ \
+
\left[
\beta_{\rm GUP}j^2
+
6\gamma Q^2 C_{(s)}
\right]
u_c
=
0.
\label{eqcriticalpoly}
\end{eqnarray}
Once the physical positive solution $r_c$ is determined, the corresponding
critical inverse temperature follows from
\begin{equation}
\tau_c
=
\tau_{(s)}(r_c).
\label{eqtauc}
\end{equation}

For the $\gamma=-1$ branch, the radial field has opposite signs at the two
ends of the physical radial domain. Consequently, within the parameter range
considered here, one positively wound defect persists and no pair-annihilation
transition occurs. The situation is different for $\gamma=+1$, for which the
field has the same sign at both boundaries. In this branch the two defects
identified in  subsection \ref{isec4B} can merge and disappear through the fold
condition of Eq.~\eqref{eqspincritical}.

For the fiducial values $Q=1$, $\Lambda=-0.1$, $\beta_{\rm GUP}=0.1$,
$m_s=1$, and $j=1$, the critical configurations of the $\gamma=+1$ branch
depend on the particle spin as summarized in Table~\ref{tabspincritical}.

\begin{table}[ht!]
\centering
\begin{tabular}{ccc}
\hline\hline
$s$ & $r_c/r_0$ & $\tau_c$
\\
\hline
$\tfrac12$ & $3.5429$ & $21.8522$
\\
$1$ & $3.5436$ & $21.1371$
\\
$\tfrac32$ & $3.5437$ & $21.0047$
\\
$2$ & $3.5437$ & $20.9584$
\\
\hline\hline
\end{tabular}
\caption{Critical radius and critical inverse temperature of the
$\gamma=+1$ branch for different particle-spin sectors. The parameters are
$Q=1$, $\lambda=2$, $\Lambda=-0.1$, $\beta_{\rm GUP}=0.1$, $m_s=1$, $j=1$,
and $r_0=1$.}
\label{tabspincritical}
\end{table}

The critical radius depends only weakly on the particle spin, whereas the
critical inverse temperature exhibits a clear systematic variation, decreasing
as $s$ increases. Thus, although the total topological charge remains $W=0$,
the boundary between the two-defect and defect-free configurations depends on
the spin of the emitted particle.

\begin{figure}[ht!]
  \centering
  \includegraphics[width=\columnwidth]{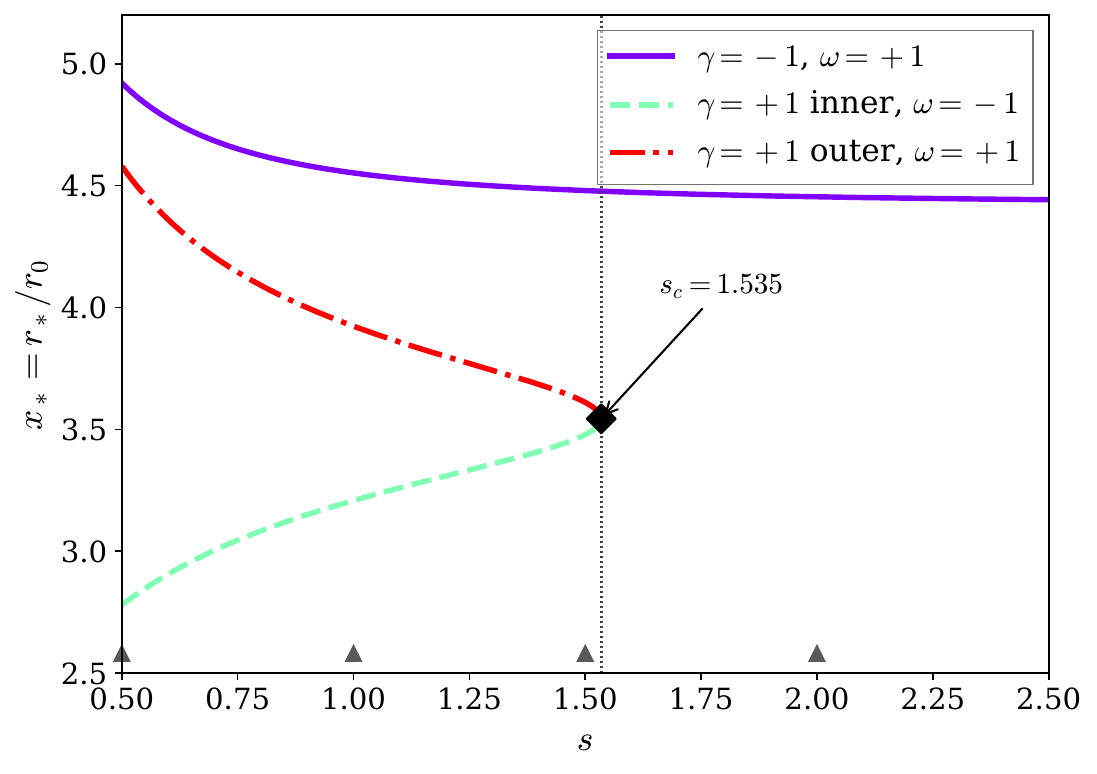}
  \caption{{\it{Defect radius $x_\ast=r_\ast/r_0$ as a function of the continuously
extended spin parameter $s$ at $\tau=21$, obtained from Eq.~\eqref{eqphirspin}. The
$\gamma=-1$ branch contains a single defect with winding number $\omega=+1$. For
$\gamma=+1$, the inner and outer defects carry $\omega=-1$ and $\omega=+1$,
respectively, and merge at $s_c\simeq1.535$ and $x_c\simeq3.544$. For $s>s_c$ no
defect remains in the $\gamma=+1$ branch.
Triangular markers indicate the physical spin values \(s=1/2,1,3/2,2\), while the
vertical dotted line marks \(s_c\simeq1.535\).  The remaining parameters are as in
Fig.~\ref{figspingm1}.}}}
  \label{figdefectvsspin}
\end{figure}

To display this behavior more clearly, we may formally extend $s$ to a
continuous control parameter for the purpose of tracing the fold. This
continuation is used only as a diagnostic tool, while the physical particle
sectors retain their discrete spin values. Fixing
$
\tau=21$
and keeping the remaining parameters at their fiducial values, the two defects
of the $\gamma=+1$ branch approach each other as $s$ increases. For the
representative physical spin sectors we obtain
\begin{equation}
\begin{aligned}
s=\frac12
\qquad&
r_{h,1}\simeq2.780\,r_0,
\qquad
r_{h,2}\simeq4.580\,r_0,
\\
s=1
\qquad&
r_{h,1}\simeq3.208\,r_0,
\qquad
r_{h,2}\simeq3.923\,r_0,
\\
s=\frac32
\qquad&
r_{h,1}\simeq3.478\,r_0,
\qquad
r_{h,2}\simeq3.611\,r_0.
\end{aligned}
\label{eqspintransitionroots}
\end{equation}
The two roots merge at
\begin{equation}
s_c\simeq1.535,
\qquad
r_c\simeq3.544\,r_0,
\label{eqcriticalspin}
\end{equation}
while for $s>s_c$ no defect remains at this value of $\tau$. Hence, the
$s=3/2$ sector lies close to the fold, whereas the $s=2$ sector is already in
the defect-free region. The proximity of $s_c$ to $3/2$ is specific to the
adopted parameter values and does not indicate a distinguished physical role
for spin-$3/2$ particles. Since the merging defects carry $\omega=-1$ and
$\omega=+1$, their annihilation leaves the total charge unchanged at $W=0$.
The evolution of the defect positions with the continuously extended spin
parameter is displayed in Fig.~\ref{figdefectvsspin}.

\begin{figure*}[ht!]
  \centering
  \includegraphics[width=0.9\textwidth]{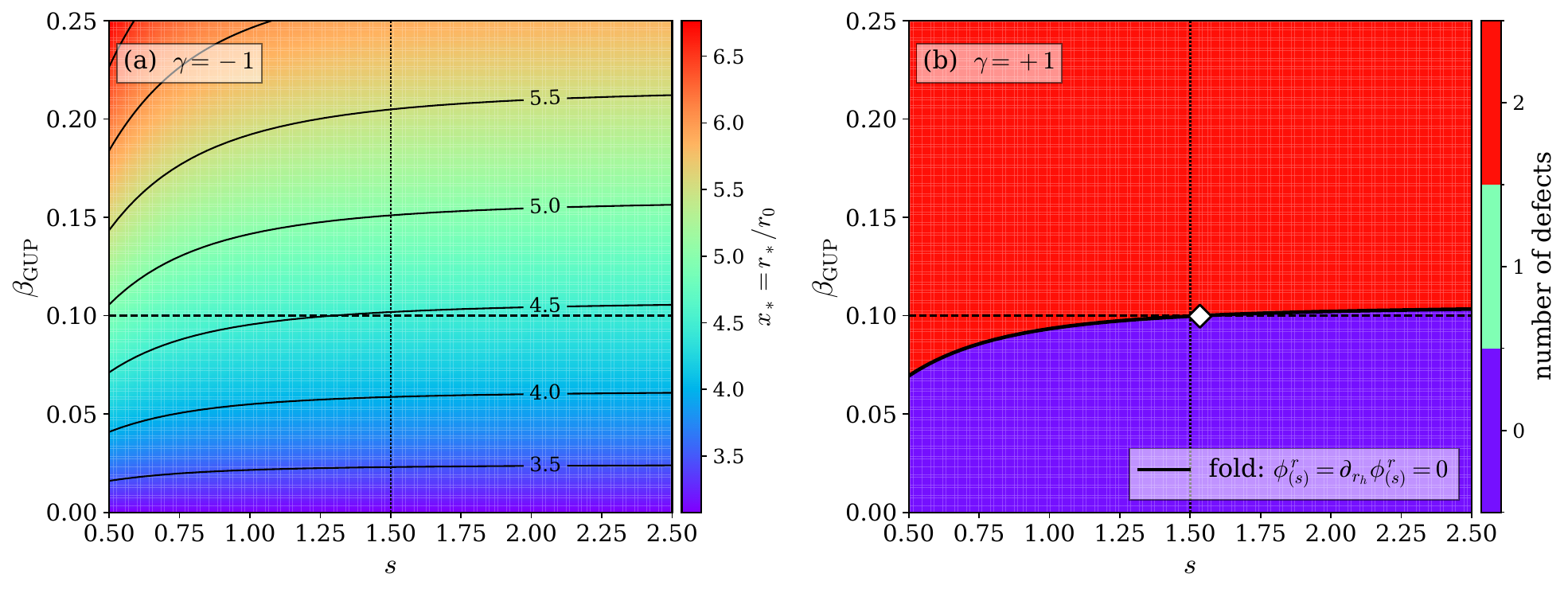}
  \caption{{\it{Particle-spin and GUP dependence of the thermodynamic defects at
$\tau=21$. For the $\gamma=-1$ branch, the left panel shows the position
$x_\ast=r_\ast/r_0$ of the single defect, which persists throughout the displayed
$(s,\beta_{\rm GUP})$ region while shifting continuously with the particle
parameters. For the $\gamma=+1$ branch, the right panel shows the number of defects,
with a defect-free region separated from a two-defect region by the fold defined by
$\phi^r_{(s)}=\partial_{r_h}\phi^r_{(s)}=0$. The two defects in the latter region
carry opposite winding numbers. The region $\beta_{\rm GUP}\gtrsim0.1$ is shown as a
formal continuation of the first-order effective description and is not used for
quantitative conclusions. Other parameters are as in
Fig.~\ref{figspingm1}.}}}
  \label{figspinbetamap}
\end{figure*}

The transition is also sensitive to the GUP parameter. For fixed spin,
increasing $\beta_{\rm GUP}$ shifts the critical inverse temperature towards
larger values. For $s=3/2$ and $\Lambda=-0.1$ we obtain
{\small{
\begin{equation}
\begin{array}{c|cccccc}
\beta_{\rm GUP}
&
0
&
0.05
&
0.10
&
0.15
&
0.20
&
0.25
\\
\hline
\tau_c
&
19.068
&
20.036
&
21.005
&
21.973
&
22.942
&
23.911
\end{array}.
\label{eqbetasweep}
\end{equation}}}
We stress that the underlying tunneling calculation is derived to first order in
$\beta_{\rm GUP}$. Hence, the values beyond the perturbative benchmark region
$\beta_{\rm GUP}\sim 0.1$ are displayed only to illustrate the continuation of the
fold structure within the first-order effective description, and should not be
interpreted quantitatively. Importantly, the critical value relevant for the
transition discussed here, $\beta_{\rm GUP}^{c}\simeq0.0998$, lies within the
perturbative region.
At fixed $\tau=21$, the defect pair is therefore absent for sufficiently small
$\beta_{\rm GUP}$ and appears once the GUP correction exceeds the critical
value
$
\beta_{\rm GUP}^{c}
\simeq0.0998$,
for $s=3/2$. Hence, the GUP parameter acts as a control parameter for the
local defect structure.
The combined dependence of the defect structure on $s$ and
$\beta_{\rm GUP}$ is displayed in Fig.~\ref{figspinbetamap}.

The cosmological constant also influences the fold. For $s=3/2$ and
$\beta_{\rm GUP}=0.1$, increasing the magnitude of the negative cosmological
constant shifts both the critical radius and the critical inverse temperature
towards smaller values. Representative results are
\begin{equation}
\begin{array}{c|cccccc}
\Lambda
&
-0.05
&
-0.08
&
-0.10
&
-0.15
&
-0.20
&
-0.30
\\
\hline
r_c/r_0
&
4.775
&
3.892
&
3.544
&
3.007
&
2.690
&
2.315
\\
\tau_c
&
30.316
&
23.667
&
21.005
&
16.850
&
14.364
&
11.412
\end{array}.
\label{eqLambdasweep}
\end{equation}
The location of the fold is therefore controlled jointly by the black-hole
parameters and by the quantities entering the GUP correction.

The limit $\beta_{\rm GUP}\rightarrow0$ provides an additional consistency
check. In this case Eq.~\eqref{eqcriticalpoly} reduces to the
Bekenstein-Hawking fold condition, and for $\gamma=+1$ one recovers
\begin{equation}
r_c\simeq3.524\,r_0,
\qquad
\tau_c\simeq19.068,
\end{equation}
in agreement with the reference analysis of subsection \ref{isec3B}. Thus, the
particle-spin dependence disappears continuously in the semiclassical limit.

These results show that the GUP correction leaves the global topological
classification of the two branches unchanged while modifying their local defect
structure and critical configurations. In particular, the particle spin shifts
the fold separating the two-defect and defect-free regions of the
$\gamma=+1$ branch, while the GUP parameter controls its location and extent.
The fold discussed here characterizes a bifurcation of the off-shell
thermodynamic-defect structure and should not by itself be identified with a
conventional equilibrium phase transition. The thermodynamic topology therefore
retains a clear dependence on the quantum properties of the emitted particle,
even though the global topological charge remains fixed.

\subsection{Consolidated topological classification}
\label{isec4D}

We can now combine the results of the reference sectors and the spin-dependent
analysis into a unified thermodynamic-topological classification.
Table~\ref{tabmaster} summarizes the defect structure obtained for the
Bekenstein-Hawking entropy, the exponentially corrected entropy, and the
GUP-corrected particle-spin sectors. For the spin-dependent case, the results
refer to $\tau=18$, with the remaining parameters fixed to the fiducial values
adopted throughout the analysis.

\begin{table}[ht!]
\centering
\begin{tabular}{lccc}
\hline\hline
sector
&
$\gamma$
&
defect structure
&
$W$
\\
\hline
Bekenstein-Hawking
&
$-1$
&
one defect, $\omega=+1$
&
$+1$
\\
Bekenstein-Hawking
&
$+1$
&
two defects, $\omega=-1,+1$
&
$0$
\\
\hline
exponentially corrected
&
$-1$
&
one defect, $\omega=+1$
&
$+1$
\\
exponentially corrected
&
$+1$
&
two defects, $\omega=-1,+1$
&
$0$
\\
\hline
spin GUP
&
$-1$
&
one defect for all $s$
&
$+1$
\\
spin GUP
&
$+1$
&
two defects for all $s$
&
$0$
\\
\hline\hline
\end{tabular}
\caption{Consolidated thermodynamic-topological classification of the
Bekenstein-Hawking, exponentially corrected, and particle-spin-dependent
sectors. The spin-GUP results refer to $s=1/2$, $1$, $3/2$, and $2$ at
$\tau=18$, with $Q=1$, $\lambda=2$, $\Lambda=-0.1$,
$\beta_{\rm GUP}=0.1$, $m_s=1$, $j=1$, and $r_0=1$.}
\label{tabmaster}
\end{table}

Several features follow directly from Table~\ref{tabmaster}. Firstly, the
$\gamma=-1$ branch belongs to the $W=+1$ class in all sectors considered
here. In the particle-spin-dependent case it contains a single positively
wound defect whose radial position changes systematically with $s$. Secondly,
the $\gamma=+1$ branch belongs to the $W=0$ class. Whenever defects are
present, they occur as a pair with opposite winding numbers and therefore
yield a vanishing total charge.

The exponentially corrected entropy preserves the same global classification
as the Bekenstein-Hawking sector for the parameter region studied here. Its
effect is concentrated predominantly at small horizon radii and is already
strongly suppressed at the defect positions relevant to the present analysis.
The GUP correction produces a more pronounced modification of the local
thermodynamic structure, since it shifts the defect positions and the critical
configurations at which oppositely wound pairs merge.

The resulting global classification can therefore be summarized as
\begin{equation}
W=
\begin{cases}
+1, & \gamma=-1,
\\
0, & \gamma=+1,
\end{cases}
\label{eqWtotalclassification}
\end{equation}
for the Bekenstein-Hawking, exponentially corrected, and
particle-spin-dependent sectors considered in this work. Importantly,
Eq.~\eqref{eqWtotalclassification} does not imply that the thermodynamic
topology is insensitive to the particle spin. Rather, the global degree
remains fixed while the local organization of the thermodynamic defects
depends on the quantum properties of the emitted particle.

This distinction is particularly clear for the $\gamma=+1$ branch. As shown
in subsection \ref{isec4C}, the particle spin shifts the critical inverse
temperature at which the pair of oppositely wound defects merges. For fixed
$\tau$, varying $s$ can therefore move the system between a two-defect
configuration and a defect-free configuration, although both belong to the
same global class $W=0$. The GUP parameter produces an analogous shift, while
the limit $\beta_{\rm GUP}\rightarrow0$ continuously recovers the
Bekenstein-Hawking result.

For completeness, it is useful to compare the resulting classification with
representative black-hole families previously studied within the
thermodynamic-topology framework. Their reported total charges are collected
in Table~\ref{tabcompare}.

\begin{table}[ht!]
\centering
\begin{tabular}{lcc}
\hline\hline
black-hole family
&
$W$
&
reference
\\
\hline
Schwarzschild
&
$-1$
&
\cite{WOS:001043962200006}
\\
Reissner-Nordstr\"om
&
$0$
&
\cite{WOS:001043962200006}
\\
Reissner-Nordstr\"om-AdS
&
$+1$
&
\cite{WOS:000964332800006}
\\
Gauss-Bonnet AdS
&
$+1$
&
\cite{WOS:000989369500001}
\\
dilatonic charged
&
$0,\pm1$
&
\cite{WOS:001362395600002}
\\
Kerr-Sen-AdS
&
$\pm1$
&
\cite{WOS:001427635500001}
\\
Born-Infeld AdS
&
$0,+1$
&
\cite{WOS:001265972500001}
\\
massive-gravity AdS $(\gamma=-1)$
&
$+1$
&
this work
\\
massive-gravity AdS $(\gamma=+1)$
&
$0$
&
this work
\\
\hline\hline
\end{tabular}
\caption{Representative total topological charges reported for different
black-hole families together with the classification obtained in the present
analysis. For the massive-gravity AdS black hole, the two values correspond
to the two branches distinguished by $\gamma$.}
\label{tabcompare}
\end{table}

The comparison shows that the global charges obtained here belong to classes
already encountered in other black-hole geometries. The distinctive feature
of the present construction lies instead in the dependence of the defect
positions and bifurcation structure on the quantum numbers of the emitted
particle. Thus, while the background geometry and thermodynamic ensemble
determine the global topological class, the GUP-deformed Hawking radiation
controls how the individual defects are organized within that class.

Hence, within the first-order GUP approximation considered here, the particle
spin does not change the total topological charge of either branch.
Nevertheless, it leaves a direct imprint on the thermodynamic-topological
structure by shifting the equilibrium defects and the critical configurations
at which oppositely wound defects merge or annihilate. This distinction
between a protected global degree and a particle-dependent local defect
structure constitutes the main topological result of the present analysis.

\section{Conclusions}
\label{Conclusions}

Black-hole thermodynamics provides a natural framework in which gravitational,
quantum, and statistical aspects of the theory meet, while thermodynamic
topology offers a complementary way of organizing black-hole equilibrium
configurations in terms of defects, winding numbers, and global topological
charges. In this work we have combined this framework with quantum-corrected
Hawking radiation in a massive-gravity AdS black-hole background. The geometry
contains two branches distinguished by $\gamma=\pm1$, which for $\lambda=2$
display qualitatively different thermodynamic behavior, with the $\gamma=-1$
branch reducing to the Reissner-Nordstr\"om-AdS form and the $\gamma=+1$
branch exhibiting a Schwarzschild-AdS-like structure. This setting therefore
provides a useful framework for examining how quantum corrections to the
radiation process are reflected in black-hole thermodynamic topology.

The main motivation was to address a question that has remained largely
unexplored. Thermodynamic-topology studies usually determine the defect
structure from the black-hole geometry, entropy, and thermodynamic ensemble,
whereas tunneling calculations allow the quantum numbers of the emitted
particle to modify the Hawking temperature. It is therefore important to
understand whether this particle dependence is lost once the thermodynamic
description is constructed, or whether it can propagate into the
thermodynamic vector field itself. In particular, we aimed to determine
whether particle spin can alter the positions and stability of thermodynamic
defects, shift the configurations at which they merge or annihilate, or even
change the global topological class. This question probes whether black-hole
thermodynamic topology is fixed entirely by the background geometry or can
also retain information about the quantum channel through which the black hole
radiates.

To investigate this issue, we combined the arbitrary-spin zitterbewegung
formalism with a generalized-uncertainty-principle deformation and obtained a
particle-spin-dependent Hawking temperature. We then used the first law to
construct the corresponding effective entropy and generalized off-shell free energy
with
an independent inverse-temperature parameter $\tau$. The resulting
thermodynamic field satisfies the required consistency condition that its zeros
reproduce $\tau=1/T_{ZB}$ to the order considered. As reference sectors, we
first analyzed the standard Bekenstein-Hawking entropy and an exponentially
corrected entropy. Both lead to the same global classification, namely
$W=+1$ for $\gamma=-1$ and $W=0$ for $\gamma=+1$, although the latter modifies
the off-shell structure at sufficiently small horizon radii.

The particle-spin-dependent sector reveals a richer local structure. For
$\gamma=-1$, one positively wound defect persists for all the physical spin
values $s=1/2$, $1$, $3/2$, and $2$, while its position shifts systematically
with the emitted-particle spin. For $\gamma=+1$, two defects with opposite
winding numbers are present at $\tau=18$, with the inner defect carrying
$\omega=-1$ and the outer one $\omega=+1$. Their positions also depend
systematically on $s$, with increasing spin bringing the two defects closer
together. Nevertheless, the asymptotic behavior of the radial thermodynamic
field is not altered by the GUP and spin-dependent terms. Consequently, the
global topological charges remain
\begin{equation}
W=
\begin{cases}
+1, & \gamma=-1,
\\
0, & \gamma=+1,
\end{cases}
\end{equation}
for all sectors investigated. Thus, the particle spin modifies the local
thermodynamic-topological structure without changing the global degree.

The particle dependence becomes particularly clear in the bifurcation
structure of the $\gamma=+1$ branch. The two oppositely wound defects merge at
a fold determined by $\phi^{r}_{(s)}=\partial_{r_h}\phi^{r}_{(s)}=0$, and the
location of this fold depends on the particle spin and on the GUP parameter.
For the fiducial parameter set, the critical inverse temperature decreases
systematically as the spin increases, while at fixed $\tau=21$ the two defects
merge at $s_c\simeq1.535$ and $r_c\simeq3.544\,r_0$. The GUP parameter acts as
an additional control parameter, with the $s=3/2$ sector at $\tau=21$
crossing the fold near $\beta_{\rm GUP}^{c}\simeq0.0998$. The cosmological
constant shifts the critical radius and inverse temperature as well. These
results show that the quantum properties of the emitted particle can reorganize
the equilibrium-defect structure and move the boundaries between different
local thermodynamic configurations, even though the configurations remain
within the same global topological class. This separation between a protected
global degree and a particle-dependent local defect structure constitutes the
main result of the present work and provides a direct connection between
quantum-corrected Hawking radiation and the corresponding effective off-shell
thermodynamic topology.

We emphasize that the particle-dependent entropy $S_{(s)}$ and the associated free
energy should be understood as channel-dependent effective thermodynamic quantities
constructed from the GUP-corrected tunneling temperature. They do not imply that the
intrinsic microscopic entropy of a fixed black hole depends on the species of the
emitted particle. Accordingly, the particle-spin dependence established here concerns
the local defect and bifurcation structure of the corresponding effective off-shell
thermodynamic description, while the underlying Bekenstein-Hawking entropy and
background geometry remain unchanged.

Several extensions could follow from this analysis. It would be
interesting to go beyond the first-order GUP approximation and determine
whether the protected global classification persists when higher-order
minimal-length corrections become important. The framework can also be
extended to other particle quantum numbers and to charged or rotating emitted
fields, where additional couplings may enrich the defect structure. Applying
the same construction to rotating black holes would allow the interplay
between the angular momentum of the background and the spin of the emitted
particle to be examined directly. Further directions include other modified
gravity backgrounds and thermodynamic ensembles, as well as a dynamical study
of the off-shell fold configurations. Such analyses may clarify how
generally quantum properties of Hawking radiation can be encoded in the local
thermodynamic topology of black holes and whether they can eventually be
connected with observable or microscopic aspects of black-hole physics. These
investigations are left for future projects.

\begin{acknowledgments}
The authors acknowledge the networking support of COST Actions CA22113
(``Fundamental challenges in theoretical physics''),
CA21106 (``COSMIC WISPers
in the Dark Universe''), CA23130 (``Bridging high and low energies in search of
quantum gravity (BridgeQG)''), CA21136 (``Addressing observational tensions in
cosmology with systematics and fundamental physics (CosmoVerse)''), CA23115
(``Relativistic Quantum Information (RQI-Action)''), and CA24101
(``Testing Fundamental Physics with Seismology (FuSe)'').
  \.I.S. and E.S. would like to thank
EMU, T\"UB\.ITAK, ANKOS, and SCOAP3 for academic support.
\end{acknowledgments}

\bibliography{finalref}

\end{document}